\documentclass[amsmath,amssymb,aps,pre,reprint, showpacs, superscriptaddress]{revtex4-1}
\usepackage{graphicx}
\usepackage{amsthm}
\usepackage{amsmath}
\usepackage{amssymb}
\usepackage{dsfont}
\usepackage{bm}
\usepackage{tikz}
\usepackage{tikz-3dplot}
\usetikzlibrary{shapes,arrows}
\usepackage{epstopdf}
\usepackage{hyperref}
\usepackage{color}

\tikzset{%
pics/.cd,
nodea/.style args={#1#2#3}{
  code={\node[minimum height=2cm] (#3) {\color{#1}#2};
       \draw[thick] (#3.south west) -| (#3.north east)--(#3.north west);
  }
},
nodeb/.style args={#1#2#3}{
  code={\node[minimum height=2cm] (#3) {\color{#1}#2};
       \draw[thick] (#3.south east) -| (#3.north west)--(#3.north east);
  }
},
nodec/.style args={#1#2#3}{
  code={\node[draw,thick,shape=circle,inner sep=1cm] (#3) {\color{#1}#2};
  }
},
}

\begin{document}
\title[]{Diverging, but negligible power at Carnot efficiency: theory and experiment}
\author{Viktor Holubec}
\email{viktor.holubec@mff.cuni.cz}
\affiliation{ 
Institut f{\"u}r Theoretische Physik, 
Universit{\"a}t Leipzig, 
Postfach 100 920, D-04009 Leipzig, Germany
}
\affiliation{ 
 Charles University,  
 Faculty of Mathematics and Physics, 
 Department of Macromolecular Physics, 
 V Hole{\v s}ovi{\v c}k{\' a}ch 2, 
 CZ-180~00~Praha, Czech Republic 
}
\author{Artem Ryabov}
\affiliation{ 
 Charles University,  
 Faculty of Mathematics and Physics, 
 Department of Macromolecular Physics, 
 V Hole{\v s}ovi{\v c}k{\' a}ch 2, 
 CZ-180~00~Praha, Czech Republic 
}
\date{\today} 
\begin{abstract} 
We discuss the possibility of reaching the Carnot efficiency by heat engines (HEs) out of quasi-static conditions at nonzero power output. We focus on several models widely used to describe the performance of actual HEs. These models comprise quantum thermoelectric devices, linear irreversible HEs, minimally nonlinear irreversible HEs, HEs working in the regime of low dissipation, over-damped stochastic HEs and an under-damped stochastic HE. Although some of these HEs can reach the Carnot efficiency at nonzero and even diverging power, the magnitude of this power is always negligible compared to the maximum power attainable in these systems.  We provide conditions for attaining the Carnot efficiency in the individual models and explain practical aspects connected with reaching the Carnot efficiency at large power output. Furthermore, we show how our findings can be tested in practice using a standard Brownian HE realizable with available micromanipulation techniques.
\end{abstract}


\maketitle

\section{Introduction}

Ever since first heat engines (HEs) appeared, engineers and physicists optimized their output power and efficiency \cite{Muller2007}. The most influential theoretical result in this field was achieved by Carnot already in the
beginning of 19th century \cite{Carnot1978}. Consider a HE which can communicate with heat baths at temperatures ranging from $T_c$ to $T_h$. Then, regardless the details of the machine, the ratio $\eta = W/Q_h$ of work done by the engine to heat accepted from surroundings is bounded from above by the Carnot efficiency $\eta_C = 1-T_c/T_h$. Recently, a lot of studies discussed whether and how this efficiency can (or can not) be actually attained in practice \cite{Benenti2011,Allahverdyan2013,Verley2014,Proesmans2015,Brandner2015,Polettini2015,Shiraishi2017, Shiraishi2016,Lee2016,Koning2016,Ponmurugan2016}.

\begin{figure}
\centerline{
  \begin{tikzpicture}[scale=0.65]
\node[draw=none, black] at (-1.5cm,1cm) {a)};
\draw[fill=red,thick] (-0.7cm,-0.7cm) rectangle (0.7cm,0.7cm);
\node[draw=none, white] at (0cm,0cm) {$T_h$};
\node[draw=none, red] at (0.4cm,-1.2cm) {$q_h$};
\draw[fill=white,thick] (0cm,-3cm) circle (1.3cm);
\node[draw=none, blue] at (0.4cm,-4.8cm) {$q_c$};
\draw[fill=blue,thick] (-0.7cm,-6.7cm) rectangle (0.7cm,-5.3cm);
\node[draw=none, white] at (0cm,-6.0cm) {$T_c$};
\draw[->,thick, black] (0cm,-0.7cm) -- (0cm,-1.7cm);
\draw[->,thick, black] (-0.0cm,-4.3cm) -- (0cm,-5.3cm);
\draw[->,thick, black] (1.3cm,-3cm) -- (2.3cm,-3cm);
\node[draw=none, black] at (1.8cm,-2.7cm) {$P$};
\node[draw=none, black] at (3.0cm,1cm) {b)};
\draw[fill=red,thick] (4.8cm,-0.7cm) rectangle (6.2cm,0.7cm);
\node[draw=none, white] at (5.5cm,0cm) {$T_h$};
\node[draw=none, red] at (6.0cm,-1.2cm) {$Q_h$};
\draw[fill=white,thick] (3.7cm,-4.3cm) rectangle (7.3cm,-1.7cm);
\draw[->,thick, red] (3.7cm,-1.7cm) -- (6.2cm,-1.7cm);
\draw[thick, red] (6.2cm,-1.7cm) -- (7.3cm,-1.7cm);
\draw[->,thick, black] (7.3cm,-1.7cm) -- (7.3cm,-3cm);
\draw[->,thick, blue] (7.3cm,-4.3cm) -- (4.8cm,-4.3cm);
\draw[thick, blue] (4.8cm,-4.3cm) -- (3.7cm,-4.3cm);
\draw[->,thick, black] (3.7cm,-4.3cm) -- (3.7cm,-3cm);
\node[draw=none, black] at (5.5cm,-3cm){$W$};
\draw[->,thick, black] (5.5cm,-4.3cm) -- (5.5cm,-5.3cm);
\node[draw=none, blue] at (6.0cm,-4.8cm) {$Q_c$};
\draw[fill=blue,thick] (4.8cm,-6.7cm) rectangle (6.2cm,-5.3cm);
\node[draw=none, white] at (5.5cm,-6.0cm) {$T_c$};
\draw[->,thick, black] (5.5cm,-0.7cm) -- (5.5cm,-1.7cm);
  \end{tikzpicture}
    }
\caption{(Color online) Two major classes of HEs which can achieve the Carnot efficiency. These HEs communicate with reservoirs at the temperatures $T_h$ and $T_c$ only. Steady-state HEs (panel a)) are coupled simultaneously to both reservoirs and operate under time-independent conditions. They transform the difference between the steady heat influx $q_h$ and the steady heat outflow $q_c$ into the output power $P$. Periodically driven HEs coupled always to a single bath at a time (panel b)) operate in a time-periodic non-equilibrium steady-state. These engines accept the total amount of heat $Q_h$ ($-Q_c$) from the hot bath (cold bath) per cycle of the duration $t_p$ and deliver the average output power $P = W/t_p$. Black branches of the cycle are adiabats, other branches are isotherms.}
\label{fig:HEs}
\end{figure}
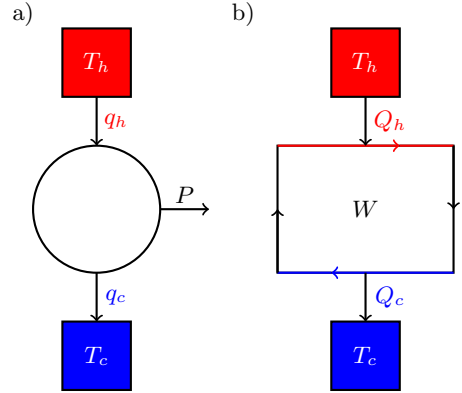

Two general classes of systems where the Carnot efficiency can be reached according to the second law are depicted in Fig.~\ref{fig:HEs}. These comprise HEs coupled simultaneously to two reservoirs at temperatures $T_h$ and $T_c$ (steady-state HEs, left) and HEs which operate periodically and are always coupled to a single bath at a time (periodic HEs, right). The Carnot efficiency can be attained also in mutations of these two classes. For example, by HEs coupled simultaneously to two baths and operated periodically \cite{Marathe2007,Basu2017}. All these machines share one important feature: They can communicate with reservoirs at the boundary temperatures $T_h$ and $T_c$ only.

Let us now focus on the periodic HEs of Fig.~\ref{fig:HEs}b). Assuming that the both baths are ideal thermodynamic reservoirs (of infinite size and infinitely fast relaxation), the second law of thermodynamics states that the total amount of entropy produced per cycle, $\Delta S_{tot}$, fulfills the inequality $\Delta S_{tot} = Q_c/T_c-Q_h/T_h \ge 0$ which leads to the relation $- Q_c/Q_h = - \Delta S_{tot} T_c/Q_h - T_c/T_h \le - T_c/T_h$. Using further the definition $W = Q_h-Q_c > 0$ for the output work of the engine, $P = W/t_p$ for the output power and $\sigma = \Delta S_{tot}/t_p$ for the average amount of entropy produced per unit time during the cycle, we obtain the following expression for the efficiency $\eta = W/Q_h$,
\begin{equation}
\eta = \frac{\eta_C}{1+T_c\sigma/P} \le \eta_C.
\label{eq:eta}
\end{equation}
The same result (only with $\sigma = q_c/T_c-q_h/T_h$) holds also for the steady-state HEs, Fig.~\ref{fig:HEs}a).

The inequality (\ref{eq:eta}) shows that the Carnot efficiency can be reached if and only if $T_c \sigma/P \to 0$. A standard example where this occurs is the quasi-static limit of infinitely slow driving ($t_p \to \infty$). Then the system is during the whole cycle in thermodynamic equilibrium and $\sigma = 0$. However, in this limit, the output power of the engine vanishes. It was suggested only recently \cite{Campisi2016,Johnson2017,Polettini2016,Koning2016,Ponmurugan2016} that there exist other ways to achieve $\eta_C$. 

First, Campisi and Fazio showed that $\eta_C$ at nonzero output power can be attained in a HE working close to a critical point \cite{Campisi2016,Koning2016, Johnson2017}. Second, the Carnot efficiency can be reached in the limit of infinitely fast dynamics \cite{Polettini2016,Lee2016}. It is important to note that both these suggestions lead to diverging heat flows through the system. In the critical HE, this is caused by the diverging heat capacity of the working fluid. For the infinitely fast dynamics, by diverging rates for processes of heat exchange with reservoirs. Such diverging energy currents represent an intuitive hallmark of devices reaching $\eta_C$ out of equilibrium. Indeed, for a non-equilibrium process one naturally assumes that $T_c \sigma>0$. Then Eq.~(\ref{eq:eta}) implies that $\eta_C$ can be reached only in the limit $P \to \infty$.

It should be noted that these suggestions for reaching the Carnot efficiency are based on idealized setups. In practice, these machines may work close to $\eta_C$, but they can never reach it as discussed in Refs.~\cite{Hondou2000, Holubec2017, Shiraishi2017,Shiraishi2016}. The critical HE proposed in Ref.~\cite{Campisi2016} exhibits diverging fluctuations of output work and power \cite{Holubec2017}. In a more general study, Shiraishi and Tajima \cite{Shiraishi2017} show that $\eta_C$ cannot be reached once finite reservoir relaxation times are taken into account (see also Refs.~\cite{Shiraishi2017a, Perarnau-Llobet2017}). 

In the present paper, we answer three basic questions concerning attainability of the Carnot efficiency out of quasi-static conditions: (i) What is the magnitude of the output power of a HE operating with $\eta_C$? (ii) For what parameters can the Carnot efficiency at nonzero power be attained in widely used models? (iii) Can an actual HE operating close to $\eta_C$ at large output power be constructed using currently available experimental techniques? 

In Sec.~\ref{ref:bounds}, we answer the first question for models where the upper bound on efficiency at given power is known. These comprise quantum thermoelectric HEs \cite{Whitney2014, Whitney2015}, linear irreversible HEs \cite{Ryabov2016, ThermoelectricZhangPRE2017}, HEs working in the regime of low-dissipation and over-damped stochastic HEs \cite{Holubec2016}, minimally nonlinear irreversible HEs \cite{Long2016} and an under-damped stochastic HE \cite{Dechant2016}. The result is quite surprising: In all these models, $\eta_C$ can be reached only at output powers which are vanishingly small as compared to the maximum power $P^\star$ attainable by the device. Reaching the Carnot efficiency thus may not be the most frequent goal in engineering practice where the magnitude of the output power often represents an important component of the figure of merit.

This allows us to generalize the well-known textbook wisdom that $\eta_C$ can be reached only at vanishingly small power $P$ to the following conjecture: The Carnot efficiency can be attained only at a vanishingly small ratio $P/P^\star$. 
First, this fraction vanishes in the quasi-static
limit ($P\to 0$). Second, it vanishes for a fixed nonzero output power $P$ and diverging maximum power $P^\star\to\infty$.

In Sec.~\ref{sec:examples} we answer the second question for the models mentioned above. The Carnot efficiency at nonzero power can be reached for reasonable parameter values just in linear response HEs, low-dissipation HEs, over-damped stochastic HEs and minimally irreversible HEs. For these models, we propose a specific scaling of parameters inspired by Ref.~\cite{Polettini2016} which can bring the efficiency arbitrarily close to $\eta_C$ at $P>0$. It turns out that this is possible both for a positive average entropy production $\sigma$ where the power at $\eta_C$ diverges and for a vanishing $\sigma$ where the power can be both finite and diverging (see Secs.~\ref{sec:LinResp} and \ref{sec:Under}). To the best of our knowledge, this is for the first time the second regime of operation has been described (see however Ref.~\cite{Lee2016} for a similar result).

The answer to our third question is given is Sec.~\ref{sec:Experiment} where we describe how our results can be tested using a stochastic Brownian HE  \cite{Martinez2017,Blickle2012, Martinez2016,Gavrilov2016,Cohen2005,Jun2012,Cohen2005a,Gavrilov2016a,
Krishnamurthy2016}. 
The HE consists of a Brownian particle diffusing in a harmonic trap with time-dependent stiffness and was already realized experimentally \cite{Blickle2012,Martinez2016,Krishnamurthy2016,Martinez2017}.  We present a realistic scaling of the stiffness under which power and efficiency of the HE increase at the same time. Moreover, in contrast to the critical HE \cite{Holubec2017, Campisi2016}, this scaling leads to bounded relative power fluctuations and hence the suggested stochastic HE can operate efficiently with a well-defined output power.

\section{Bounds on maximum efficiency at given power}
\label{ref:bounds}

\begin{figure}
	\centering
		\includegraphics[width=0.9\columnwidth]{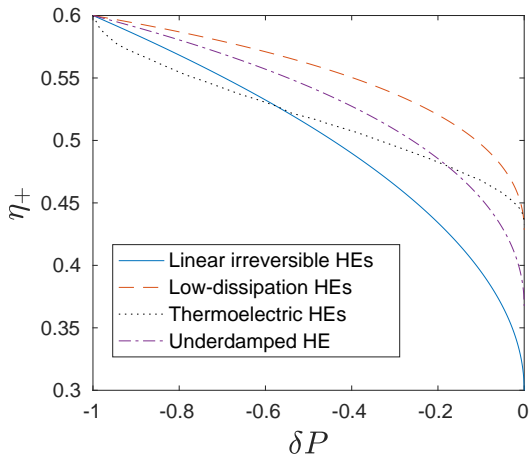}
	\caption{(Color online) Upper bound on the efficiency at given power for quantum thermoelectric HEs (dotted yellow line), for linear irreversible HEs (\ref{eq:LinRes}) (full blue line), for low-dissipation HEs and minimally nonlinear irreversible HEs (\ref{eq:LowDiss}) (dashed orange line) and for HEs based on an uderdamped Brownian particle in a breathing parabola potential (\ref{eq:Lutz}) (dot-dashed purple line); $\eta_C = 3/5$.}
	\label{fig:fig1}
\end{figure}

Upper bounds on efficiency at given power obtained in the studies \cite{Whitney2014, Whitney2015,Ryabov2016,Holubec2016,Dechant2016,Long2016} can be effectively written using the variable \cite{HolubecRyabov2015,Holubec2016a}
\begin{equation}
\delta P = \frac{P-P^\star}{P^\star},\quad \delta P \in [-1,0],
\label{eq:dP}
\end{equation}
which measures relative change in power $P$ with respect to the maximum power $P^\star$ achievable in a given setup.

Let us denote the maximum efficiency attainable at given power as $\eta_+(\delta P)$. The individual upper bounds for all the models mentioned in the introduction are plotted in Fig.~\ref{fig:fig1}. All the curves monotonously increase with decreasing $\delta P$ from the efficiency at maximum power ($\delta P = 0$) to the Carnot efficiency ($\delta P = -1$). In all the models, $\eta_C$ can thus be reached under the condition $P/P^\star \to 0$ only. Hence we have $\eta = \eta_C + f(P/P^\star)$, $\lim_{x\to +0}f(x) = 0$. For the models considered here, the function $f(x)$ is given by the power law $f(x) = c x^{\theta}$ with a negative constant $c$  ($\eta \le \eta_C$) and a positive exponent $\theta$. Close to $\eta_C$, it follows from Eq.~(\ref{eq:eta}) that $\eta = \eta_C - \eta_C \sigma T_c/P$ and we thus have
\begin{equation}
\eta-\eta_C \approx- \eta_C \frac{\sigma T_c}{P}  =  f\left(\frac{P}{P^\star}\right) = c \left(\frac{P}{P^\star}\right)^{\theta} \le 0.
\label{eq:PstarCloseToEtaC}
\end{equation}

In case the maximum power $P^\star$ is finite, $\eta_C$ can be reached only at vanishingly small power $P$ and thus at quasi-static conditions. The possibility to attain the Carnot efficiency at nonzero or even diverging power opens only when the maximum power diverges faster than $P$.

This suggests that reaching $\eta_C$ at $P>0$ may not be the holy grail of engineering practice where a trade-off between power and efficiency is often optimized \cite{Gordon1992, Baldasaro2001, Smith2004, DelCampo2014, Long2015, Haseli2013, Gonzalez-AyalaPRE2017, WangEntropy2017}. Consider for example the target function $\xi = \eta^\alpha P^\beta$, $\alpha\ge 0$, $\beta \ge 0$, which should be optimized. If we denote as $\xi_C$ and $\xi^\star$ the values of this function at maximum efficiency and at maximum power, respectively, we find that the condition $P_C/P^\star = 0$ yields $\xi_C/\xi^\star = 0$ whenever $\beta \neq 0$. Our findings thus encourage the struggle for reaching $\eta_C$ only if the ultimate performance goal is the maximum efficiency (for example at a fixed value of output power). 
In such a case, the advantage of HEs working close to $\eta_C$ under non-equilibrium conditions is their typically large power output (which is still negligible as compared to $P^{\star}$). A practical disadvantage resides in preparation of a working medium and/or operational cycle for such engines, which must be tuned in a special way.   

Having described the general formulation of our findings, let us now turn to particular examples of HEs. Below, we review the bounds obtained for the individual model HEs and discuss the conditions under which these HEs can achieve the Carnot efficiency. In Sec.~\ref{sec:thermoelectic} we discuss in short thermoelectric HEs and Sec.~\ref{sec:LinResp} contains an extensive discussion of linear response HEs. In Sec.~\ref{sec:Over} we discuss in detail low-dissipation HEs in general and Sec.~\ref{sec:Under} is devoted to a HE based on an under-damped particle diffusing in an externally controlled parabolic potential. 
Finally, in Sec.~\ref{sec:Experiment}, we propose a specific setup where $\eta_C$ at $P>0$ can be reached in experiments with Brownian particles diffusing in an externally controlled potential. We give a detailed model study including all experimentally relevant parameters.

From a practical point of view, the scalings presented in Secs.~\ref{sec:LinResp}, \ref{sec:Over} and \ref{sec:Experiment} must be understood as recipes how to set model parameters in order to attain efficiencies close to $\eta_C$ at large output power
as it is presented in Figs.~\ref{fig:fig3} and \ref{fig:fig4}. For $\eta=\eta_C$, the parameters in the individual models either diverge (Onsager coefficients for linear response HEs) or vanish (the cycle duration for low-dissipation HEs and for the Brownian HE). Such extreme values are experimentally inaccessible and brake validity of basic assumptions underlying the individual models.

\section{Examples}
\label{sec:examples}

\subsection{Quantum thermoelectric heat engines}
\label{sec:thermoelectic}

Thermoelectric HEs are connected to two thermodynamic reservoirs at different chemical potentials and temperatures (see Ref.~\cite{Benenti2017} for the latest review). They are operating in a non-equilibrium steady-state (Fig.~\ref{fig:HEs}a)) using the temperature gradient for pumping electrons against the gradient of chemical potential. For quantum thermoelectric HEs operating under vanishing magnetic fields, the upper bound on efficiency can be written analytically only for $\delta P \to -1$, where it reads \cite{Whitney2014, Whitney2015} 
\begin{equation}
\eta_+(\delta P) = \eta_C\left(
1 - 0.478 \sqrt{(1-\eta_C)(1+\delta P)}
\right).
\label{eq:TE}
\end{equation}
For the remaining values of $\delta P$ the curve can be obtained numerically \cite{Whitney2014, Whitney2015}. The resulting curve depicted in Fig.~\ref{fig:fig1} always exhibits the general features described above. 

This bound was derived by Whitney \cite{Whitney2014, Whitney2015} using nonlinear Landauer-B{\"u}ttiker scattering theory and it is generally valid for all systems which can be modeled by this theory. In particular, the bound may not be valid once the time-reversal symmetry of the underlying dynamics is broken, for example by introducing magnetic fields into the system.

The maximum power corresponding to the bound (\ref{eq:TE}) is given by $P^\star = A_0 \pi^2 N k_B^2/h (T_h-T_c)^2$, where $A_0 \pi^2 k_B^2/h$ is a constant and $N$ is the number of transverse modes in the narrowest part of the quantum system. Assuming that $N$ is finite, the bound (\ref{eq:TE}) suggests that $\eta_C$ at $P>0$ may be reached in the unrealistic case of diverging temperature difference, $T_h-T_c \to \infty$, only. 

When the efficiency  of the thermoelectric reaches $\eta_C$, the entropy production scales as $\sigma \propto P^{3/2}$ \cite{Whitney2014, Whitney2015}. The ratio $\sigma/P$ thus scales as $\sigma/P \propto \sqrt{P}$ leading to the exponent $\theta = 1/2$ in Eq.~(\ref{eq:PstarCloseToEtaC}). This can be checked by the direct calculation from Eq.~(\ref{eq:TE}).

\subsection{Linear irreversible heat engines}
\label{sec:LinResp}

Let us now focus on HEs working in the linear response regime. The linear response formalism is valid for an arbitrary system if (thermodynamic) forces acting on it are small enough that the system operates close to thermal equilibrium. The formalism can be applied both to HEs operating in a non-equilibrium steady-state caused by their simultaneous coupling to two (or more) reservoirs at different temperatures (Fig.~\ref{fig:HEs}a)) and to cyclic HEs, which are connected only to a single bath at a time (Fig.~\ref{fig:HEs}b)). Because the two classes of models where shown to be equivalent \cite{Raz2016,Proesmans2015}, we will limit our discussion to steady-state models only.

Assuming that the Onsager matrix describing the linear model is symmetric, the upper bound on efficiency at given power is given by \cite{Ryabov2016}
\begin{equation}
\eta_+(\delta P) = \frac{\eta_C}{2}\left(1+\sqrt{-\delta P}\right).
\label{eq:LinRes}
\end{equation}
For non-symmetric Onsager matrices, the Carnot efficiency can be in principle reached even at maximum power \cite{Benenti2011} in the limit $L_{12}/L_{21} \to \infty$ if one considers just the restriction imposed by the second law. However, for example for thermoelectric HEs a detailed analysis of Onsager coefficients strongly suggests that the power vanishes at least linearly when $\eta$ reaches $\eta_C$ \cite{Brandner2015a}. This result narrows the way to a thermoelectric working in the linear response regime at $\eta = \eta_C$ and $P>0$. Nevertheless, such a working regime may be still attainable even for a symmetric Onsager matrix based on considerations of Ref.~\cite{Polettini2016}.

We demonstrate this possibility on the simple model comprising just two thermodynamic forces $X_1$, $X_2$ and fluxes $J_1$, $J_2$ related by a symmetric Onsager matrix: 
\begin{eqnarray}
J_1 &=& L_{11} X_1 + L_{12} X_2,
\label{eq:J1}\\
J_2 &=& L_{12} X_1 + L_{22} X_2.
\label{eq:J2}
\end{eqnarray}
The first thermodynamic force $X_1 = F/T$ determines the output power $P = - J_1 X_1 T$ ($F$ is the load attached to the engine and $T$ is the system temperature). The second thermodynamic force $X_2 = (T_h-T_c)/T^2$ impels the heat flux $J_2$ from the hot reservoir to the system. The engine efficiency is thus defined as $\eta = P/J_2$.

The maximum power $P^\star = \eta_C^2 L_{22} q^2 T/4$ is in this system attained for the load $X_1^\star = - L_{12} X_2/(2 L_{11})$ at the efficiency $\eta^\star = 0.5 \eta_C q^2/(2-q^2)$ \cite{BroeckPRL2005, Ryabov2016}. The constant $q^2 = L_{12}^2/(L_{11}L_{22})$, characterizes coupling between the fluxes $J_1$ and $J_2$, which become proportional for $q^2 =1$. The definition of the entropy production in the system, $\sigma = J_1 X_1 + J_2 X_2$, together with the second law, $\sigma > 0$, implies the following limitations for the Onsager coefficients: $L_{11} \ge 0$, $L_{22} \ge 0$ and $L_{11}L_{22} - L_{12}^2 \ge 0$. These restrictions give the bounds $-1 \le q \le 1$ for the coupling constant $q$.

With these definitions and results, the maximum efficiency at given power can be written as \cite{Ryabov2016}
\begin{equation}
\eta = \eta^\star (1 + \delta P) \frac{2-q^2}{2-q^2(1 + \sqrt{-\delta P})}.
\end{equation}
Optimization of this result with respect to $q^2$ gives the formula (\ref{eq:LinRes}) and the optimal value of the parameter $q=1$. To sum up, $\eta_C$ can be in this model attained only when $q = 1$ and $\delta P = -1$.

The ratio $P/P^\star$ can be written as $P/P^\star = (2-X_1/X_1^\star)X_1/X_1^\star$. Thus the condition $\delta P = -1$ implies either $X_1 = 0$ or $X_1/X_1^\star = - (2L_{11}X_1)/(L_{12}X_2) = 2$, and from $q^2 = 1$ it follows that $L_{12}/L_{11} = L_{22}/L_{12}$. When both these conditions are strictly satisfied, the output power $P$ as well as the entropy production $\sigma$ vanish, meaning that $\eta_C$ is reached under quasi-static conditions. 

However, the way is not completely closed yet, as suggested in Refs.~\cite{Polettini2015a, Polettini2016}. This is because we can set $q^2=1$ and ensure that $\delta P$ approaches $-1$ asymptotically such that neither $P$ nor $\sigma$ vanish and, at the same time, $\eta$ approaches $\eta_C$. Setting $L_{12}/L_{11} = L_{22}/L_{12}$ ($q^2=1$), output power and entropy production can be written as
\begin{eqnarray}
P &=& - L_{12}X_1 X_2 \left( 1 + \frac{L_{11}X_1}{L_{12}X_2}\right)T,
\label{eq:P_LR}
\\
\sigma &=& L_{12}X_1 X_2 \left( \frac{L_{11}X_1}{L_{12}X_2} + 2 +  \frac{L_{12}X_2}{L_{11}X_1}\right).
\label{eq:sigma_LR}
\end{eqnarray} 

For $q^2=1$, only two free parameters remain, e.g.\ $L_{11}$ and $L_{12}$. We now let them diverge and assume that thermodynamic forces $X_1$ and $X_2$ are bounded, which is reasonable in the linear response regime. Instead of working directly with the two Onsager coefficients, it is convenient to split the flux $J_1$ such that $L_{12} X_2 = (J_{\infty} + J_1)$ and  $L_{11} X_1 = -J_{\infty}$. Assuming that $J_{\infty} \gg J_1 > 0$, the condition $- L_{11}/L_{12} = X_2/X_1$ (or $\delta P =-1$) is strictly obeyed in the limit $J_{\infty} \to \infty$ only. Inserting these expressions into Eqs.~(\ref{eq:P_LR})-(\ref{eq:sigma_LR}), we determine asymptotic behavior of $\sigma$, $P$ and $P^\star$ as $J_{\infty} \to \infty$. We obtain $\sigma \approx - X_1 J_1^2/J_{\infty}$,  $P \approx X_2 J_1$ and $\sigma/P \approx J_1/J_{\infty}$ for their ratio which must vanish as $\eta \to \eta_C$. The ratio $P/P^\star$ reads
${P}/{P^\star} \approx 4 J_1/(T^2 J_\infty)$ and hence it vanishes in the same way as $\sigma T/P$. The exponent $\theta$ in Eq.~(\ref{eq:PstarCloseToEtaC}) thus equals to $1$.

The Carnot efficiency at nonzero power can thus be reached for $J_{\infty} \to \infty$. Let us assume that $J_1 \propto J_\infty^\kappa$, $\kappa < 1$. Then the power diverges as $J_{\infty} \to \infty$ whenever $\kappa > 0$ and is finite otherwise. In the same limit, the entropy production $\sigma$ vanishes whenever $\kappa < 1/2$, is constant for $\kappa = 1/2$ and diverges otherwise. By a proper choice of Onsager coefficients, one can hence increase the output power, reach efficiencies arbitrarily close to $\eta_C$ and, at the same time, suppress the entropy production.

To conclude, the Carnot efficiency can be reached by properly tuning parameters of the working medium encoded in Onsager coefficients. To get close to $\eta_C$, these coefficients must attain relatively large values. In practice, large Onsager coefficients can be obtained for binary mixtures near their critical point (see Ref.~\cite{Binder2007} and the references therein). Our results thus provide further evidence that $\eta_C$ at $P>0$ can be attained by HEs working close to a critical point as suggested in Ref.~\cite{Campisi2016}.

\subsection{Low-dissipation and Minimally nonlinear irreversible heat engines}
\label{sec:Over}

For low-dissipation HEs \cite{Holubec2016} and for minimally nonlinear irreversible HEs \cite{Long2016} the upper bound on efficiency at given power reads
\begin{equation}
\eta_+(\delta P) = \eta_C \frac{1+\sqrt{-\delta P}}{2 - \left(1-\sqrt{-\delta P}\right)\eta_C}. 
\label{eq:LowDiss}
\end{equation}
The minimally nonlinear irreversible HEs represent straightforward generalization of linear irreversible HEs discussed in the preceding subsection. Within this generalization, it is assumed that the formulas (\ref{eq:J1})--(\ref{eq:J2}) for the currents also contain quadratic terms of the type $\kappa_i J_i^2$. The setup used to derive the bound (\ref{eq:LowDiss}) for minimally nonlinear irreversible HEs \cite{Long2016} is mathematically equivalent to the low-dissipation model \cite{Izumida2012}. The latter is discussed in next paragraphs including a straightforward physical interpretation of its basic assumptions.

\begin{figure}
	\centering
		\includegraphics[width=0.9\columnwidth]{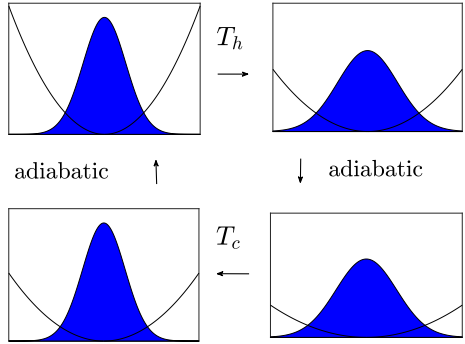}
	\caption{(Color online) Sketch of the operational cycle of HEs based on a particle diffusing in a time-dependent potential considered in Sections \ref{sec:Over}, \ref{sec:Under} and \ref{sec:Experiment}. The filled blue curve stands for the probability density for the position of the particle. The black parabola represents the potential at the beginning of the individual branches.}
	\label{fig:fig2}
\end{figure}

The low-dissipation model used to derive the bound~(\ref{eq:LowDiss}) is depicted schematically in Fig.~\ref{fig:HEs}b) and it can be mapped onto the Brownian HE depicted in Fig.~\ref{fig:fig2}, see Refs.~\cite{HolubecRyabov2015,Holubec2016a,Esposito2010b,Schmiedl2008, Holubec2014, Holubec2016}. In Fig.~\ref{fig:fig2}, the filled blue Gaussian represents the probability density function (PDF) for the position of a Brownian particle driven by a time-dependent potential (black line). During the cycle, the system is first attached for time $t_h$  to the hot bath and the potential widens (the isothermal expansion). Then the bath temperature changes from $T_h$ to $T_c$ and the potential further opens. We assume that this happens so fast
that the PDF remains unchanged (the adiabatic step). After that, the system is attached for time $t_c$ to the cold bath and the potential shrinks (the isothermal compression). Finally, the non-equilibrium Carnot cycle is closed as the temperature and the potential jump to their initial values (the adiabatic step). 

The system performs work during the first two strokes when the potential opens and consumes work in the rest of the cycle. The cycle was already realized experimentally using a colloidal particle manipulated by optical tweezers by Blickle and Bechinger \cite{Blickle2012}.

The main assumption of the low-dissipation model \cite{Esposito2010b,Gonzalez-AyalaEntropy2017, Johal2017PRE} is that the heat exchanged with the individual reservoirs during the isotherms can be written as 
\begin{eqnarray}
Q_h &=& T_h \Delta S - A_h/t_h,
\label{eq:Qh}\\
Q_c &=& T_c \Delta S + A_c/t_c,
\label{eq:Qc}
\end{eqnarray}
leading to the entropy produced per cycle
\begin{equation}
\Delta S_{tot} = \frac{A_h}{t_hT_h} + \frac{A_c}{t_cT_c}.
\label{eq:Stot}
\end{equation} 
Here, $A_h$ and $A_c$ are positive parameters independent of the times $t_{h}$ and $t_c$ and $\Delta S$ is the change of the system entropy during the isothermal expansion. Low-dissipation HEs thus work at vanishing entropy production under quasi-static conditions ($t_h\to \infty$ and $t_c \to \infty$). 

Although the assumptions (\ref{eq:Qh})--(\ref{eq:Qc}) may look oversimplified, there exist several real systems which fit into the scheme. Examples are nanomotors based on two-level quantum systems \cite{Zulkowski2015a} and various over-damped Brownian HEs  \cite{Schmiedl2008, Zulkowski2015, HolubecRyabov2015,Holubec2016a,Martinez2016}. In general, jumps in the temperature at the ends of the isotherms bring the system far from equilibrium leading to additional terms in the total entropy production (\ref{eq:Stot}) which would not vanish in the long-time limit. To avoid this, one should drive the system in such a way that it is in equilibrium both before and after the jump. While this can be relatively easily achieved for large systems, it might require a precise control of the system dynamics on the micro-scale \cite{HolubecRyabov2015,Holubec2016a,Schmiedl2008, Holubec2014,Sato2002,Zulkowski2015a,
Zulkowski2015,Martinez2016}.

Let us now investigate if low-dissipation HEs can operate with the Carnot efficiency at nonzero power. In Sec.~\ref{sec:Experiment}, we will exemplify the obtained results using an exactly solvable Brownian HE \cite{Holubec2014,HolubecRyabov2015,Holubec2016a,Schmiedl2008} which can be realized experimentally \cite{Blickle2012, Martinez2016,Martinez2017,Krishnamurthy2016}.

Because the adiabatic branches are (infinitely) faster than the isotherms, the total duration of the cycle is given by $t_p \approx t_h + t_c$. Introducing the parameter $\alpha$ through the formulas $t_h = \alpha t_p$ and $t_c = (1-\alpha) t_p$, the power of a low-dissipation HE is given by \cite{Schmiedl2008}
\begin{equation}
P = \frac{(T_h - T_c)\Delta S}{t_p} - \frac{(1-\alpha)A_h +  \alpha A_c}{t_p^2\alpha(1-\alpha)}.
\label{eq:power_Wirr}
\end{equation} 
Optimization of the power with respect to $t_p$ and $\alpha$ gives
\begin{eqnarray}
\alpha^{\star} &=& \frac{A_h - \sqrt{A_hA_c}}{A_h - A_c},
\label{eq:alpha_opt}\\
t_p^{\star} &=& \frac{2}{T_h\eta_C\Delta S}(\sqrt{A_h}+\sqrt{A_c})^2,
\label{eq:tp_opt}\\
P^{\star} &=& \frac{1}{4}\left(\frac{T_h\eta_C\Delta S}{\sqrt{A_h} + \sqrt{A_c}}\right)^2,
\label{eq:P_opt}\\
\eta^{\star} &=& \frac{\eta_C(1+\sqrt{A_c/A_h})}{2(1+\sqrt{A_c/A_h})  - \eta_C}
\label{eq:eta_opt}.
\end{eqnarray}
The upper bound for efficiency at maximum power, $\eta_+(0) =\eta_ C/(2-\eta_C)$, and also the bound~(\ref{eq:LowDiss}) are obtained in the limit $A_c/A_h \to 0$. Nevertheless, qualitatively similar bounds as (\ref{eq:LowDiss}) apply for arbitrary $A_c$ and $A_h$ (see Fig.~2b in Ref.~\cite{Holubec2016}).

The Carnot efficiency at nonzero power can be attained only if the maximum power diverges, which occurs for $(\sqrt{A_h}+\sqrt{A_c})^2 \to 0$. To approach this condition asymptotically, we assume that the coefficients scale as $A_c \propto A_\infty^{-\phi}$ and $A_h \propto A_\infty^{-\delta}$,  $A_\infty \to \infty$. Then both $A_c/A_h \to 0$ and $(\sqrt{A_h}+\sqrt{A_c})^2 \to 0$ whenever $0<\delta<\phi$. The condition $(\sqrt{A_h}+\sqrt{A_c})^2 \to 0$ alone is fulfilled for $0<\phi\le\delta$.

Vanishing coefficients $A_c$ and $A_h$ lead to a vanishing entropy production (\ref{eq:Stot}) unless the cycle duration $t_p$ also goes to zero. This occurs in the regime of maximum power, where $t_p$ (\ref{eq:tp_opt}) scales as $t_p^\star \propto A_h$. Larger efficiencies than $\eta^\star$ are obtained for $t_p > t_p^\star$ \cite{Holubec2016}. Let us thus assume that $t_p = t_p^\star A_h^{-\kappa}$, $\kappa > 0$ and $\alpha = \alpha^\star$. 

Using this scaling we obtain the relation $P \approx (T_h\eta_C\Delta S)^2 A_h^{\kappa - 1}/2$ for power and the formula ${\sigma}/P = A_h^{\kappa}/(2 T_h)$ for the ratio which must vanish at $\eta=\eta_C$. The power at the Carnot efficiency is constant for $\kappa = 1$ and diverges for $0 < \kappa < 1$. The entropy produced per cycle scales as $\Delta S_{tot} \approx 2 A_h^{\kappa}$ and the average entropy produced per unit time is given by $\sigma = T_h(\eta_C\Delta S)^2 A_h^{2\kappa - 1}/2$. We hence see that the average entropy produced per cycle vanishes whenever $\eta=\eta_C$ at $P>0$. The entropy produced per unit time vanishes for $\kappa<1/2$, is nonzero for $\kappa = 1/2$ and diverges otherwise. We obtain the same surprising result as in the linear response: By a proper choice of parameters one can increase the power, achieve efficiencies arbitrary close to $\eta_C$ and, at the same time, suppress the entropy production.

Finally, the ratio $P/P^\star$ reads $P/P^\star \approx 2 A_h^{\kappa}$ and thus it scales in the same way as $\Delta S_{tot}$ and ${\sigma}/P$ (exponent $\theta$ in Eq.~(\ref{eq:PstarCloseToEtaC}) equals to $1$).

The discussed setting represents another example of reaching $\eta_C$ at $P>0$ by fast driving as suggested in Ref.~\cite{Polettini2016} for a different model. In Section \ref{sec:Experiment}, we will show that a HE based on an over-damped Brownian particle driven by optimally controlled parabolic potential is a low dissipation HE. Interestingly enough, this is not the case for an under-damped particle.

\subsection{Under-damped Brownian heat engine}
\label{sec:Under}

For a stochastic HE based on an under-damped Brownian particle driven by optimally controlled parabolic potential the bound on efficiency at fixed power is \cite{Dechant2016}
\begin{multline}
\eta_+(\delta P) =
\eta_{CA}(1+\delta P) - \frac{\eta_C}{2}\delta P + \\
+\frac{1}{2} \sqrt{-\delta P} \sqrt{\eta_C^2 - \eta_{CA}^4 (1+\delta P)},
\label{eq:Lutz}
\end{multline}
where $\eta_{CA} = 1- \sqrt{T_c/T_h}$ stands for the famous Curzon-Ahlborn efficiency \cite{Curzon1975,Chambadal1957,Novikov1958,Yvon1955}. The model used for the derivation is based on the cycle depicted in Fig.~\ref{fig:fig2}.

The maximum power is given by $P^\star = \gamma_h \gamma_c \left(\sqrt{T_h} - \sqrt{T_c}\right)^2/\left(\sqrt{\gamma_h}+ \sqrt{\gamma_c}\right)^2$
and thus it diverges for diverging temperature differences or for diverging friction constants $\gamma_h$ and $\gamma_c$, in complete contradiction to what is found in the over-damped limit (see Eqs.~(\ref{eq:P_opt}) and (\ref{eq:AhcOver})). The former regime of operation is unphysical and the latter one breaks the basic assumption of the model. The formula (\ref{eq:Lutz}) has been derived assuming small friction coefficient as compared to the frequency $\omega$ of the parabolic potential $\omega x^2$. Under this assumption, $\eta_C$ can be reached only for very strong potentials. 

From Eqs.~(\ref{eq:PstarCloseToEtaC}) and (\ref{eq:Lutz}) it follows $\sigma/P = \eta_{CA}^2/(T_h \eta_C) P/P^\star$ and thus the exponent $\theta$ in Eq.~(\ref{eq:PstarCloseToEtaC})
is for the present model equal to $1$.

\section{Proposed experiment: \\
Brownian heat engine}
\label{sec:Experiment}

Brownian HEs are frequently used to demonstrate and verify latest results in stochastic thermodynamics \cite{HolubecRyabov2015,Holubec2016a,Schmiedl2008,Blickle2012, Martinez2017,Blickle2012, Martinez2016,Gavrilov2016,Cohen2005,Jun2012,Cohen2005a,Gavrilov2016a}. In this section, we show how to tune control parameters of a realistic over-damped HE such that it works close to $\eta_C$ at large output power.

We consider the one-dimensional Brownian HE with the working cycle depicted in Fig.~\ref{fig:fig2}. The probability density for the particle position satisfies the Fokker-Planck equation \cite{Risken1996,Schmiedl2008}
\begin{align}
\frac{\partial}{\partial t}p(x,t) &= - \frac{\partial}{\partial x} j(x,t)\,,
\label{eq:Fokker-Planck}\\
j(x,t) &= - \frac{1}{\gamma(t)}\left\{ k_{B}T(t)\frac{\partial}{\partial x} +  \left[\frac{\partial U(x,t)}{\partial x}\right]  \right\}p(x,t)\,,
\label{eq:current}
\end{align}
supplemented by the periodicity condition $p(x,t + t_p) = p(x,t)$. Above, $k_B$ is the Boltzmann constant, $\gamma(t)$ denotes the friction constant, $T(t)$ stands for the actual temperature of the reservoir coupled to the particle and $U(x,t)$ denotes the externally controlled potential. The friction constant depends on temperature. Thus we have $T(t) = T_h$, $\gamma(t) = \gamma_h$ along the hot isotherm and $T(t) = T_c$, $\gamma(t) = \gamma_c$ along the cold one.

Thermodynamics of the Brownian HE is described by Eqs.~(\ref{eq:Qh}) and (\ref{eq:Qc}) with the parameters
\cite{Schmiedl2008} (see also Eqs. (8) and (10) in Ref.~\cite{Pigolotti2017})
\begin{equation}
A_{h,c} = \gamma_{h,c} t_{h,c} \int_{t_i^{h,c}}^{t_f^{h,c}}dt
\int_{-\infty}^{\infty}dx \frac{j^2(x,t)}{p(x,t)}.
\end{equation}
Here $t_i^h$ ($t_f^h$) denote the initial (final) time of the hot isotherm of the cycle and $t_i^c$ ($t_f^c$) denote the same for the cold one. In general, the parameters $A_{h,c}$ depend on durations of the two isotherms in a nontrivial way. However, once the time dependence of the potential is optimized to yield maximum output work, the parameters $A_{h,c}$ become independent of $t_{h,c}$. Then the Brownian HE is a low dissipation HE.

Further, we assume the specific potential 
\begin{equation}
U(x,t) = \frac{k(t)}{2}x^2,
\label{eq:potential}
\end{equation}
for which the optimization procedure can be performed analytically \cite{Schmiedl2008,Holubec2014} and which is easily created by optical tweezers \cite{Blickle2012,Martinez2016,Martinez2017,Krishnamurthy2016}. The resulting optimal driving $k(t)$
contains a discontinuity connected with the instantaneous change of temperature during the adiabatic branches \cite{Holubec2014}. 
Let us denote as $k_h(t)$ ($k_c(t)$) the optimal protocol during the hot (cold) isotherm. These functions read
\begin{eqnarray}
k_h(t) &=& \frac{1}{w_0}\frac{k_B T_h}{(1+b_1 t)^2} - \frac{\gamma_h b_1}{1+b_1 t},\\
k_c(t) &=& \frac{1}{w_f}\frac{k_B T_c}{[1+b_2 (t-t_h)]^2} - \frac{\gamma_c b_2}{1+b_2 (t-t_h)}.
\label{eq:driving_max_power}
\end{eqnarray}
Here, the parameter $w_0$ ($w_f$) stands for the variance of the particle position at the beginning (end) of the hot isotherm (see Fig.~\ref{fig:fig2}). The constants $b_1$ and $b_2$ are given by $b_1 = \left(\sqrt{w_f/w_0} - 1\right)/t_h$ and $b_2 = \left(\sqrt{w_0/w_f} - 1\right)/t_c$ \footnote{Although the presented thermodynamic analysis of the Brownian HE is standard, it neglects heat currents connected with momentum degrees of freedom \cite{Schmiedl2008,Martinez2015, Arold2017}. These heat currents inevitably lower the efficiency. In the over-damped limit, the momentum is assumed to be in equilibrium and the effect of the corresponding heat flow on the efficiency can be canceled out by introducing a regenerator. If this is not possible, one should use different optimal protocols than those derived in the over-damped limit, especially with the aim to facilitate the additional heat flux into the momentum space. Such optimization procedure cannot be performed analytically and is out of the scope of the present paper.}. 

Using this driving, the parameters $A_h$ and $A_c$ are given by
\begin{equation}
A_{h,c} = \gamma_{h,c} \left(\sqrt{w_f} - \sqrt{w_0} \right)^2.
\label{eq:AhcOver}
\end{equation}
The scaling $A_c \propto A_\infty^{-\phi}$ and $A_h \propto A_\infty^{-\delta}$ proposed in Sec.~\ref{sec:Over} to reach the Carnot efficiency in the limit $A_\infty \to \infty$ thus implies that $\eta_C$ can be reached either for vanishing friction constants $\gamma_{h,c}$ or for the vanishing bracket $\left(\sqrt{w_f} - \sqrt{w_0} \right)$. The assumption of small friction constants contradicts conditions of the over-damped limit. The ratio $w_f/w_0$ determines the increase in system entropy during the hot isotherm, $\Delta S = k_B\log\left(w_f/w_0\right)/2$, and thus the reversible work done by the system. In order to achieve small parameters $A_{h,c}$ we thus assume that the particle is during the whole cycle strongly localized, i.e. $w_{0,f} \to 0$, while we keep constant $w_f/w_0$. 

\begin{figure}
	\centering
		\includegraphics[width=1.0\columnwidth]{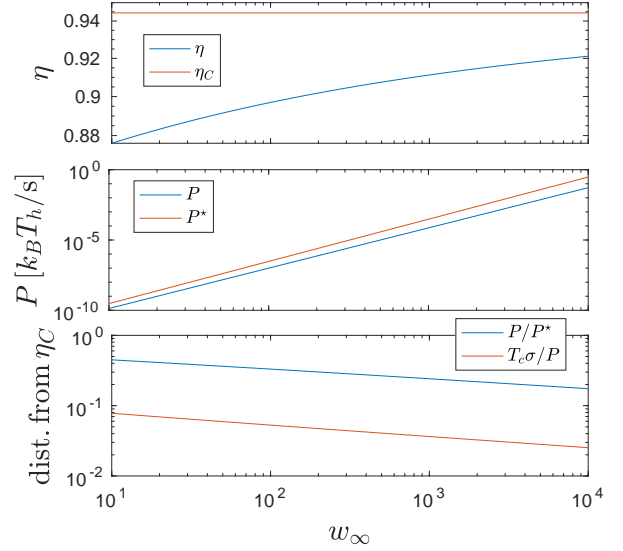}
	\caption{(Color online) Efficiency $\eta$ (upper panel), power $P$ and maximum power $P^\star$ (middle panel) and two variables (\ref{eq:PstarCloseToEtaC}) measuring the distance from the Carnot efficiency of the engine described in Sec.~\ref{sec:Experiment} as functions of the scaling parameter $w_{\infty}$. 
The increasing efficiency is accompanied with the increasing power in direct contradiction with the quasi-static limit. The detailed description and the parameters used are given in Sec.~\ref{sec:Experiment}.}
	\label{fig:fig3}
\end{figure}

In micromanipulation experiments with Brownian particles, the Carnot efficiency can be achieved as follows. We set $\delta = \phi$ and $\gamma_h = \gamma_c$ and thus $A_h = A_c$. The assumption of equal friction coefficients is realistic for changing the temperature in accord with the study \cite{Martinez2016}. We also take equal durations of the two isothermal branches, $t_h = t_c$ (thus $\alpha = 1/2$). Finally, we assume that the largest variance during the cycle scales as $w_f = w_\infty^{-\xi}$, $\xi>0$, and thus the coefficients $A_{h,c}$ are given by $A_h = A_c \propto w_\infty^{-\xi}$, i.e. $\delta = \phi = \xi$.

\begin{figure}
	\centering
		\includegraphics[width=1.0\columnwidth]{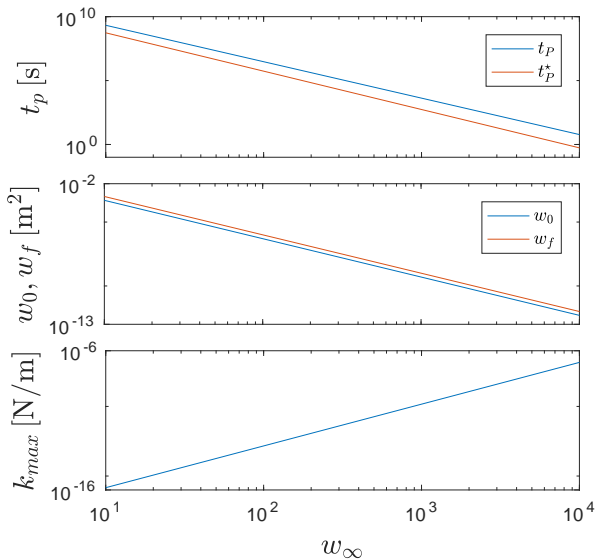}
	\caption{(Color online) Total cycle duration $t_p$ and duration of the cycle at maximum power $t_p^\star$ (upper panel), minimum and maximum variances of the particle position during the cycle $w_0$ and $w_f$ (middle panel) and the maximum value of the trap stiffness (lower panel) for the engine described in Sec.~\ref{sec:Experiment} as functions of the scaling parameter $w_{\infty}$. The region accessible in experiments is roughly $w_\infty \in (10^3, 10^4)$ where the cycle time decreases from one hour to 5 seconds. The detailed description and the parameters used are given in Sec.~\ref{sec:Experiment}.}
	\label{fig:fig4}
\end{figure}

\begin{figure}
	\centering
		\includegraphics[width=0.8\columnwidth]{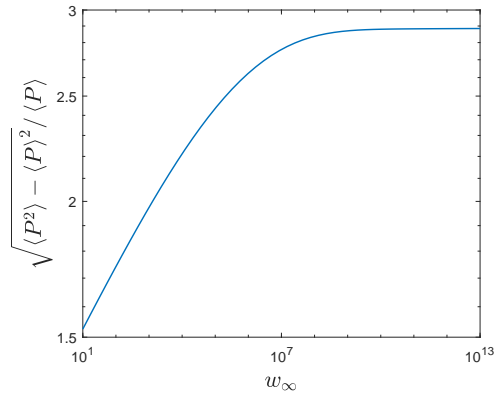}
	\caption{(Color online) The relative fluctuation of power for the Brownian HE described in Sec.~\ref{sec:Experiment} as a function of the scaling parameter $w_{\infty}$. The detailed description and the parameters used are given in Sec.~\ref{sec:Experiment}.}
	\label{fig:fig5}
\end{figure}

In the numerical illustration shown in Figs.~\ref{fig:fig3}--\ref{fig:fig5} we consider a very light colloidal particle of the diameter $R = 10^{-6}$~m diffusing in water with the friction coefficient given by the Stokes' law: $\gamma_h=\gamma_c=\gamma = 6\pi R \mu$. Here $\mu = 1.002\times 10^{-3}$~Pa~s is the dynamic viscosity of water at the room temperature $293.15$ K. We assume that the real bath temperature during the both branches is $T_c = 293.15$ K and that during the hot isotherm this temperature is effectively increased by an additional external noise to $T_h = 5273.15$ K similarly as in the recent experimental work \cite{Martinez2016}. The corresponding Carnot efficiency is $\eta_C \approx 0.945$. We use the exponent $\kappa = 0.05$ for the cycle time $t_p$ and $\xi = -3$ for the maximum variance $w_f$, i.e. we take $t_p = t_p^\star A_h^{-0.05}$ and $w_f = w_\infty^{-3}$. Finally, we fix the ratio of the maximum and minimum variance to be $w_f/w_0 = 2$. According to Sec.~\ref{sec:Over}, this choice leads to the following scaling of the thermodynamic variables in question: $P \propto w_\infty^{2.85} $, $P^\star \propto w_\infty^{3}$, $t_p \propto w_\infty^{-2.85}$, $t_p^\star \propto w_\infty^{-3}$, $\eta_C - \eta \propto P/P^\star \propto \sigma/P \propto \Delta S_{tot} \propto  w_\infty^{-0.15}$ and $\sigma \propto w_\infty^{2.7}$.

In Fig.~\ref{fig:fig3} we show the behavior of thermodynamic variables of the system with increasing parameter $w_\infty$. The engine efficiency $\eta$ converges to $\eta_C \approx 0.945$ (upper panel). In contrast to the quasi-static limit, this increase in $\eta$ is accompanied with an increase in power (middle panel). The lower panel shows the convergence of the ratio of power to maximum power, $P/P^*$, and of the product $T_c \sigma/P$ to zero as $\eta \to \eta_C$. The two lines are parallel as predicted by Eq.~(\ref{eq:PstarCloseToEtaC}) for $\eta$ close to $\eta_C$.  

In Fig.~\ref{fig:fig4} we show experimentally controlled variables as functions of the parameter $w_\infty$. Both the cycle duration $t_p$ and the optimal cycle duration $t_p^\star$ goes from experimentally unaccessible values (years) for small $w_\infty$ to reasonable values (seconds) for large $w_\infty$. Similarly, the minimum and maximum particle variance (middle panel) are very large for small $w_\infty$ and attain realistic values for large $w_\infty$. The corresponding maximum spring constant $k_{max} = k_h(0)$ is plotted in the lower panel. The whole range of the spring constant shown in the figure, especially the part for large $w_\infty$ can be readily achieved in experiments either using optical tweezers \cite{Blickle2012, Martinez2016,Martinez2017} or a feedback or anti-Brownian electrokinetic trap \cite{Gavrilov2016,Cohen2005,Jun2012,Cohen2005a,Gavrilov2016a}.

In Fig.~\ref{fig:fig5} we show the relative fluctuation of power, $\sqrt{\left< P^2\right> - \left< P\right>^2}/P$. The calculation has been performed numerically using the procedure described in Ref.~\cite{Holubec2014}, Sec.~3.2. The power fluctuation increases with increasing scaling parameter $w_\infty$ and saturates at a relatively small value in the limit $w_\infty \to \infty$. In contrast to the critical HE introduced in Ref.~\cite{Campisi2016} (see Ref.~\cite{Holubec2017}), the proposed Brownian HE delivers a relatively stable output power.

\section{Conclusion and outlook}

The struggle to reach the Carnot efficiency at nonzero output power is an exciting part of the current research in non-equilibrium thermodynamics. In the present work, we have added another piece into the mosaic: For various models widely used to describe the performance of actual HEs, the output power at which $\eta_C$ can be possibly reached is doomed to be negligibly small as compared to the maximum power achievable in these models. This is best visible from Eq.~(\ref{eq:PstarCloseToEtaC}) which shows that both the ratio of entropy production to the output power, $\sigma/P$, and the ratio of output power to the maximum power, $P/P^\star$, must vanish when the Carnot limit is attained. 

Besides that, we have investigated conditions for reaching divergent maximum power and thus $\eta_C$ at $P>0$ in the individual models. These settings seem to be unrealistic for thermoelectric HEs (infinite temperature gradient) and for an under-damped Brownian HE (the required conditions break assumptions of the model). More realistic conditions were found for linear response HEs, minimally nonlinear irreversible HEs and low-dissipation HEs.

In the linear response regime, $\eta_C$ may be attained for diverging Onsager coefficients as also suggested in Ref.~\cite{Polettini2016}. The open question is whether such conditions can be achieved in a real system. A suitable candidate is the critical HE proposed by Campisi and Fazio \cite{Campisi2016}. 

In the low-dissipation regime, $\eta_C$ can be achieved for fast cycles with vanishing dissipation coefficients $A_h$ and $A_c$. 
In practice, these conditions can be (nearly) satisfied by Brownian HEs which might be constructed using current micro-manipulation experimental techniques. Concrete parameters and the driving protocol for a Brownian HE operating near $\eta_C$ at large $P$
are discussed in Sec.~\ref{sec:Experiment}. We believe that this detailed analysis will stimulate experimental verification of our findings. In connection with the experiment, it would be very interesting to investigate behavior of probability densities for work, heat and fluctuating efficiency \cite{Verley2014} for $\eta$ close to $\eta_C$ at large power.

At a first glance, the condition $\sigma/P \to 0$ implies that reaching $\eta_C$ at $P>0$ should require diverging power $P$, because one naturally assumes that $\sigma>0$. For linear response HEs and low-dissipation HEs, we have found certain scalings which allow reaching $\eta_C$ at $P>0$ and vanishing entropy production at the same time. The corresponding power can both diverge and attain a finite non-zero value. To the best of our knowledge, such scenario has not been discussed in previous works. In practice, this allows constructing a HE operating close to the Carnot efficiency at a large output power and with a small entropy production.

Sensitivity of the individual models to the precise form of the scaling which must be chosen in order to achieve $\eta_C$ at $P>0$ represents the biggest qualitative difference between the present approach and the quasi-static limit. In order to realize a quasi-static cycle, it is enough to make it very slow regardless the details of the system. On the contrary, the scalings leading to the Carnot efficiency at nonzero power must be engineered in a model dependent manner and, moreover, their practical usage requires precise control of the system dynamics.

Our present knowledge suggests that the limit $\eta \to \eta_C$ always incurs negative effects. In the quasi-static limit, the power at $\eta_C$ vanishes. For critical HEs \cite{Campisi2016,Johnson2017, Holubec2017}, approaching $\eta_C$ at $P>0$ is accompanied by diverging power fluctuations (even in the macroscopic limit). For HEs studied in this work we observe a large loss in power as compared to the maximum power regime. Interestingly, such negative effects are not present in machines working under isothermal conditions in a steady-state driven by chemical or external forces \cite{Seifert2011}. Some of these machines can reach the second law upper bound on efficiency under maximum power conditions.

\begin{acknowledgments}
Support of the work by the Czech Science Foundation (project No. 17-06716S) is gratefully acknowledged. VH in addition gratefully acknowledges the support by Alexander von Humboldt Foundation.
\end{acknowledgments}

%



\begin{thebibliography}{73}%
\makeatletter
\providecommand \@ifxundefined [1]{%
 \@ifx{#1\undefined}
}%
\providecommand \@ifnum [1]{%
 \ifnum #1\expandafter \@firstoftwo
 \else \expandafter \@secondoftwo
 \fi
}%
\providecommand \@ifx [1]{%
 \ifx #1\expandafter \@firstoftwo
 \else \expandafter \@secondoftwo
 \fi
}%
\providecommand \natexlab [1]{#1}%
\providecommand \enquote  [1]{``#1''}%
\providecommand \bibnamefont  [1]{#1}%
\providecommand \bibfnamefont [1]{#1}%
\providecommand \citenamefont [1]{#1}%
\providecommand \href@noop [0]{\@secondoftwo}%
\providecommand \href [0]{\begingroup \@sanitize@url \@href}%
\providecommand \@href[1]{\@@startlink{#1}\@@href}%
\providecommand \@@href[1]{\endgroup#1\@@endlink}%
\providecommand \@sanitize@url [0]{\catcode `\\12\catcode `\$12\catcode
  `\&12\catcode `\#12\catcode `\^12\catcode `\_12\catcode `\%12\relax}%
\providecommand \@@startlink[1]{}%
\providecommand \@@endlink[0]{}%
\providecommand \url  [0]{\begingroup\@sanitize@url \@url }%
\providecommand \@url [1]{\endgroup\@href {#1}{\urlprefix }}%
\providecommand \urlprefix  [0]{URL }%
\providecommand \Eprint [0]{\href }%
\providecommand \doibase [0]{http://dx.doi.org/}%
\providecommand \selectlanguage [0]{\@gobble}%
\providecommand \bibinfo  [0]{\@secondoftwo}%
\providecommand \bibfield  [0]{\@secondoftwo}%
\providecommand \translation [1]{[#1]}%
\providecommand \BibitemOpen [0]{}%
\providecommand \bibitemStop [0]{}%
\providecommand \bibitemNoStop [0]{.\EOS\space}%
\providecommand \EOS [0]{\spacefactor3000\relax}%
\providecommand \BibitemShut  [1]{\csname bibitem#1\endcsname}%
\let\auto@bib@innerbib\@empty
\bibitem [{\citenamefont {M{\"u}ller}(2007)}]{Muller2007}%
  \BibitemOpen
  \bibfield  {author} {\bibinfo {author} {\bibfnamefont {I.}~\bibnamefont
  {M{\"u}ller}},\ }\href {https://books.google.de/books?id=u13KiGlz2zcC} {\emph
  {\bibinfo {title} {A History of Thermodynamics: The Doctrine of Energy and
  Entropy}}}\ (\bibinfo  {publisher} {Springer},\ \bibinfo {year}
  {2007})\BibitemShut {NoStop}%
\bibitem [{\citenamefont {Carnot}(1978)}]{Carnot1978}%
  \BibitemOpen
  \bibfield  {author} {\bibinfo {author} {\bibfnamefont {S.}~\bibnamefont
  {Carnot}},\ }\href {https://books.google.cz/books?id=J3JzXJxYZfIC} {\emph
  {\bibinfo {title} {R{\'e}flexions sur la puissance motrice du feu}}},\ edited
  by\ \bibinfo {editor} {\bibfnamefont {R.}~\bibnamefont {Fox}},\ Acad{\'e}mie
  internationale d'histoire des sciences. Collection des travaux\ (\bibinfo
  {publisher} {J. Vrin},\ \bibinfo {year} {1978})\BibitemShut {NoStop}%
\bibitem [{\citenamefont {Benenti}\ \emph {et~al.}(2011)\citenamefont
  {Benenti}, \citenamefont {Saito},\ and\ \citenamefont
  {Casati}}]{Benenti2011}%
  \BibitemOpen
  \bibfield  {author} {\bibinfo {author} {\bibfnamefont {G.}~\bibnamefont
  {Benenti}}, \bibinfo {author} {\bibfnamefont {K.}~\bibnamefont {Saito}}, \
  and\ \bibinfo {author} {\bibfnamefont {G.}~\bibnamefont {Casati}},\ }\href
  {\doibase 10.1103/PhysRevLett.106.230602} {\bibfield  {journal} {\bibinfo
  {journal} {Phys. Rev. Lett.}\ }\textbf {\bibinfo {volume} {106}},\ \bibinfo
  {pages} {230602} (\bibinfo {year} {2011})}\BibitemShut {NoStop}%
\bibitem [{\citenamefont {Allahverdyan}\ \emph {et~al.}(2013)\citenamefont
  {Allahverdyan}, \citenamefont {Hovhannisyan}, \citenamefont {Melkikh},\ and\
  \citenamefont {Gevorkian}}]{Allahverdyan2013}%
  \BibitemOpen
  \bibfield  {author} {\bibinfo {author} {\bibfnamefont {A.~E.}\ \bibnamefont
  {Allahverdyan}}, \bibinfo {author} {\bibfnamefont {K.~V.}\ \bibnamefont
  {Hovhannisyan}}, \bibinfo {author} {\bibfnamefont {A.~V.}\ \bibnamefont
  {Melkikh}}, \ and\ \bibinfo {author} {\bibfnamefont {S.~G.}\ \bibnamefont
  {Gevorkian}},\ }\href {\doibase 10.1103/PhysRevLett.111.050601} {\bibfield
  {journal} {\bibinfo  {journal} {Phys. Rev. Lett.}\ }\textbf {\bibinfo
  {volume} {111}},\ \bibinfo {pages} {050601} (\bibinfo {year}
  {2013})}\BibitemShut {NoStop}%
\bibitem [{\citenamefont {Verley}\ \emph {et~al.}(2014)\citenamefont {Verley},
  \citenamefont {Esposito}, \citenamefont {Willaert},\ and\ \citenamefont
  {Van~den Broeck}}]{Verley2014}%
  \BibitemOpen
  \bibfield  {author} {\bibinfo {author} {\bibfnamefont {G.}~\bibnamefont
  {Verley}}, \bibinfo {author} {\bibfnamefont {M.}~\bibnamefont {Esposito}},
  \bibinfo {author} {\bibfnamefont {T.}~\bibnamefont {Willaert}}, \ and\
  \bibinfo {author} {\bibfnamefont {C.}~\bibnamefont {Van~den Broeck}},\ }\href
  {https://www.nature.com/articles/ncomms5721} {\bibfield  {journal} {\bibinfo
  {journal} {Nat. Commun.}\ }\textbf {\bibinfo {volume} {5}},\ \bibinfo {pages}
  {4721} (\bibinfo {year} {2014})}\BibitemShut {NoStop}%
\bibitem [{\citenamefont {Proesmans}\ and\ \citenamefont {Van~den
  Broeck}(2015)}]{Proesmans2015}%
  \BibitemOpen
  \bibfield  {author} {\bibinfo {author} {\bibfnamefont {K.}~\bibnamefont
  {Proesmans}}\ and\ \bibinfo {author} {\bibfnamefont {C.}~\bibnamefont
  {Van~den Broeck}},\ }\href {\doibase 10.1103/PhysRevLett.115.090601}
  {\bibfield  {journal} {\bibinfo  {journal} {Phys. Rev. Lett.}\ }\textbf
  {\bibinfo {volume} {115}},\ \bibinfo {pages} {090601} (\bibinfo {year}
  {2015})}\BibitemShut {NoStop}%
\bibitem [{\citenamefont {Brandner}\ \emph {et~al.}(2015)\citenamefont
  {Brandner}, \citenamefont {Saito},\ and\ \citenamefont
  {Seifert}}]{Brandner2015}%
  \BibitemOpen
  \bibfield  {author} {\bibinfo {author} {\bibfnamefont {K.}~\bibnamefont
  {Brandner}}, \bibinfo {author} {\bibfnamefont {K.}~\bibnamefont {Saito}}, \
  and\ \bibinfo {author} {\bibfnamefont {U.}~\bibnamefont {Seifert}},\ }\href
  {\doibase 10.1103/PhysRevX.5.031019} {\bibfield  {journal} {\bibinfo
  {journal} {Phys. Rev. X}\ }\textbf {\bibinfo {volume} {5}},\ \bibinfo {pages}
  {031019} (\bibinfo {year} {2015})}\BibitemShut {NoStop}%
\bibitem [{\citenamefont {Polettini}\ \emph
  {et~al.}(2015{\natexlab{a}})\citenamefont {Polettini}, \citenamefont
  {Verley},\ and\ \citenamefont {Esposito}}]{Polettini2015}%
  \BibitemOpen
  \bibfield  {author} {\bibinfo {author} {\bibfnamefont {M.}~\bibnamefont
  {Polettini}}, \bibinfo {author} {\bibfnamefont {G.}~\bibnamefont {Verley}}, \
  and\ \bibinfo {author} {\bibfnamefont {M.}~\bibnamefont {Esposito}},\ }\href
  {\doibase 10.1103/PhysRevLett.114.050601} {\bibfield  {journal} {\bibinfo
  {journal} {Phys. Rev. Lett.}\ }\textbf {\bibinfo {volume} {114}},\ \bibinfo
  {pages} {050601} (\bibinfo {year} {2015}{\natexlab{a}})}\BibitemShut
  {NoStop}%
\bibitem [{\citenamefont {Shiraishi}\ and\ \citenamefont
  {Tajima}(2017)}]{Shiraishi2017}%
  \BibitemOpen
  \bibfield  {author} {\bibinfo {author} {\bibfnamefont {N.}~\bibnamefont
  {Shiraishi}}\ and\ \bibinfo {author} {\bibfnamefont {H.}~\bibnamefont
  {Tajima}},\ }\href {\doibase 10.1103/PhysRevE.96.022138} {\bibfield
  {journal} {\bibinfo  {journal} {Phys. Rev. E}\ }\textbf {\bibinfo {volume}
  {96}},\ \bibinfo {pages} {022138} (\bibinfo {year} {2017})}\BibitemShut
  {NoStop}%
\bibitem [{\citenamefont {Shiraishi}\ \emph {et~al.}(2016)\citenamefont
  {Shiraishi}, \citenamefont {Saito},\ and\ \citenamefont
  {Tasaki}}]{Shiraishi2016}%
  \BibitemOpen
  \bibfield  {author} {\bibinfo {author} {\bibfnamefont {N.}~\bibnamefont
  {Shiraishi}}, \bibinfo {author} {\bibfnamefont {K.}~\bibnamefont {Saito}}, \
  and\ \bibinfo {author} {\bibfnamefont {H.}~\bibnamefont {Tasaki}},\ }\href
  {\doibase 10.1103/PhysRevLett.117.190601} {\bibfield  {journal} {\bibinfo
  {journal} {Phys. Rev. Lett.}\ }\textbf {\bibinfo {volume} {117}},\ \bibinfo
  {pages} {190601} (\bibinfo {year} {2016})}\BibitemShut {NoStop}%
\bibitem [{\citenamefont {Lee}\ and\ \citenamefont {Park}(2017)}]{Lee2016}%
  \BibitemOpen
  \bibfield  {author} {\bibinfo {author} {\bibfnamefont {J.~S.}\ \bibnamefont
  {Lee}}\ and\ \bibinfo {author} {\bibfnamefont {H.}~\bibnamefont {Park}},\
  }\href {\doibase 10.1038/s41598-017-10664-9} {\bibfield  {journal} {\bibinfo
  {journal} {Sci. Rep.}\ }\textbf {\bibinfo {volume} {7}},\ \bibinfo {pages}
  {10725} (\bibinfo {year} {2017})}\BibitemShut {NoStop}%
\bibitem [{\citenamefont {Koning}\ and\ \citenamefont
  {Indekeu}(2016)}]{Koning2016}%
  \BibitemOpen
  \bibfield  {author} {\bibinfo {author} {\bibfnamefont {J.}~\bibnamefont
  {Koning}}\ and\ \bibinfo {author} {\bibfnamefont {J.~O.}\ \bibnamefont
  {Indekeu}},\ }\href {\doibase 10.1140/epjb/e2016-70297-9} {\bibfield
  {journal} {\bibinfo  {journal} {Eur. Phys. J. B}\ }\textbf {\bibinfo {volume}
  {89}},\ \bibinfo {pages} {248} (\bibinfo {year} {2016})}\BibitemShut
  {NoStop}%
\bibitem [{\citenamefont {{Ponmurugan}}(2016)}]{Ponmurugan2016}%
  \BibitemOpen
  \bibfield  {author} {\bibinfo {author} {\bibfnamefont {M.}~\bibnamefont
  {{Ponmurugan}}},\ }\href {https://arxiv.org/abs/1604.01912} {\bibfield
  {journal} {\bibinfo  {journal} {ArXiv e-prints}\ } (\bibinfo {year}
  {2016})},\ \Eprint {http://arxiv.org/abs/1604.01912} {arXiv:1604.01912}
  \BibitemShut {NoStop}%
\bibitem [{\citenamefont {Marathe}\ \emph {et~al.}(2007)\citenamefont
  {Marathe}, \citenamefont {Jayannavar},\ and\ \citenamefont
  {Dhar}}]{Marathe2007}%
  \BibitemOpen
  \bibfield  {author} {\bibinfo {author} {\bibfnamefont {R.}~\bibnamefont
  {Marathe}}, \bibinfo {author} {\bibfnamefont {A.~M.}\ \bibnamefont
  {Jayannavar}}, \ and\ \bibinfo {author} {\bibfnamefont {A.}~\bibnamefont
  {Dhar}},\ }\href {\doibase 10.1103/PhysRevE.75.030103} {\bibfield  {journal}
  {\bibinfo  {journal} {Phys. Rev. E}\ }\textbf {\bibinfo {volume} {75}},\
  \bibinfo {pages} {030103} (\bibinfo {year} {2007})}\BibitemShut {NoStop}%
\bibitem [{\citenamefont {Basu}\ \emph {et~al.}(2017)\citenamefont {Basu},
  \citenamefont {Nandi}, \citenamefont {Jayannavar},\ and\ \citenamefont
  {Marathe}}]{Basu2017}%
  \BibitemOpen
  \bibfield  {author} {\bibinfo {author} {\bibfnamefont {D.}~\bibnamefont
  {Basu}}, \bibinfo {author} {\bibfnamefont {J.}~\bibnamefont {Nandi}},
  \bibinfo {author} {\bibfnamefont {A.~M.}\ \bibnamefont {Jayannavar}}, \ and\
  \bibinfo {author} {\bibfnamefont {R.}~\bibnamefont {Marathe}},\ }\href
  {\doibase 10.1103/PhysRevE.95.052123} {\bibfield  {journal} {\bibinfo
  {journal} {Phys. Rev. E}\ }\textbf {\bibinfo {volume} {95}},\ \bibinfo
  {pages} {052123} (\bibinfo {year} {2017})}\BibitemShut {NoStop}%
\bibitem [{\citenamefont {Campisi}\ and\ \citenamefont
  {Fazio}(2016)}]{Campisi2016}%
  \BibitemOpen
  \bibfield  {author} {\bibinfo {author} {\bibfnamefont {M.}~\bibnamefont
  {Campisi}}\ and\ \bibinfo {author} {\bibfnamefont {R.}~\bibnamefont
  {Fazio}},\ }\href {\doibase doi:10.1038/ncomms11895} {\bibfield  {journal}
  {\bibinfo  {journal} {Nat. Commun.}\ }\textbf {\bibinfo {volume} {7}},\
  \bibinfo {pages} {11895} (\bibinfo {year} {2016})}\BibitemShut {NoStop}%
\bibitem [{\citenamefont {Johnson}(2017)}]{Johnson2017}%
  \BibitemOpen
  \bibfield  {author} {\bibinfo {author} {\bibfnamefont {C.~V.}\ \bibnamefont
  {Johnson}},\ }\href {https://arxiv.org/abs/1703.06119} {\bibfield  {journal}
  {\bibinfo  {journal} {arXiv:1703.06119}\ } (\bibinfo {year}
  {2017})}\BibitemShut {NoStop}%
\bibitem [{\citenamefont {Polettini}\ and\ \citenamefont
  {Esposito}(2017)}]{Polettini2016}%
  \BibitemOpen
  \bibfield  {author} {\bibinfo {author} {\bibfnamefont {M.}~\bibnamefont
  {Polettini}}\ and\ \bibinfo {author} {\bibfnamefont {M.}~\bibnamefont
  {Esposito}},\ }\href {http://stacks.iop.org/0295-5075/118/i=4/a=40003}
  {\bibfield  {journal} {\bibinfo  {journal} {EPL}\ }\textbf {\bibinfo {volume}
  {118}},\ \bibinfo {pages} {40003} (\bibinfo {year} {2017})}\BibitemShut
  {NoStop}%
\bibitem [{\citenamefont {Hondou}\ and\ \citenamefont
  {Sekimoto}(2000)}]{Hondou2000}%
  \BibitemOpen
  \bibfield  {author} {\bibinfo {author} {\bibfnamefont {T.}~\bibnamefont
  {Hondou}}\ and\ \bibinfo {author} {\bibfnamefont {K.}~\bibnamefont
  {Sekimoto}},\ }\href {\doibase 10.1103/PhysRevE.62.6021} {\bibfield
  {journal} {\bibinfo  {journal} {Phys. Rev. E}\ }\textbf {\bibinfo {volume}
  {62}},\ \bibinfo {pages} {6021} (\bibinfo {year} {2000})}\BibitemShut
  {NoStop}%
\bibitem [{\citenamefont {Holubec}\ and\ \citenamefont
  {Ryabov}(2017)}]{Holubec2017}%
  \BibitemOpen
  \bibfield  {author} {\bibinfo {author} {\bibfnamefont {V.}~\bibnamefont
  {Holubec}}\ and\ \bibinfo {author} {\bibfnamefont {A.}~\bibnamefont
  {Ryabov}},\ }\href {\doibase 10.1103/PhysRevE.96.030102} {\bibfield
  {journal} {\bibinfo  {journal} {Phys. Rev. E}\ }\textbf {\bibinfo {volume}
  {96}},\ \bibinfo {pages} {030102} (\bibinfo {year} {2017})}\BibitemShut
  {NoStop}%
\bibitem [{\citenamefont {Shiraishi}(2017)}]{Shiraishi2017a}%
  \BibitemOpen
  \bibfield  {author} {\bibinfo {author} {\bibfnamefont {N.}~\bibnamefont
  {Shiraishi}},\ }\href {\doibase 10.1103/PhysRevE.95.052128} {\bibfield
  {journal} {\bibinfo  {journal} {Phys. Rev. E}\ }\textbf {\bibinfo {volume}
  {95}},\ \bibinfo {pages} {052128} (\bibinfo {year} {2017})}\BibitemShut
  {NoStop}%
\bibitem [{\citenamefont {Perarnau-Llobet}\ \emph {et~al.}(2017)\citenamefont
  {Perarnau-Llobet}, \citenamefont {Wilming}, \citenamefont {Riera},
  \citenamefont {Gallego},\ and\ \citenamefont {Eisert}}]{Perarnau-Llobet2017}%
  \BibitemOpen
  \bibfield  {author} {\bibinfo {author} {\bibfnamefont {M.}~\bibnamefont
  {Perarnau-Llobet}}, \bibinfo {author} {\bibfnamefont {H.}~\bibnamefont
  {Wilming}}, \bibinfo {author} {\bibfnamefont {A.}~\bibnamefont {Riera}},
  \bibinfo {author} {\bibfnamefont {R.}~\bibnamefont {Gallego}}, \ and\
  \bibinfo {author} {\bibfnamefont {J.}~\bibnamefont {Eisert}},\ }\href@noop {}
  {\bibfield  {journal} {\bibinfo  {journal} {arXiv:1704.05864}\ } (\bibinfo
  {year} {2017})}\BibitemShut {NoStop}%
\bibitem [{\citenamefont {Whitney}(2014)}]{Whitney2014}%
  \BibitemOpen
  \bibfield  {author} {\bibinfo {author} {\bibfnamefont {R.~S.}\ \bibnamefont
  {Whitney}},\ }\href {\doibase 10.1103/PhysRevLett.112.130601} {\bibfield
  {journal} {\bibinfo  {journal} {Phys. Rev. Lett.}\ }\textbf {\bibinfo
  {volume} {112}},\ \bibinfo {pages} {130601} (\bibinfo {year}
  {2014})}\BibitemShut {NoStop}%
\bibitem [{\citenamefont {Whitney}(2015)}]{Whitney2015}%
  \BibitemOpen
  \bibfield  {author} {\bibinfo {author} {\bibfnamefont {R.~S.}\ \bibnamefont
  {Whitney}},\ }\href {\doibase 10.1103/PhysRevB.91.115425} {\bibfield
  {journal} {\bibinfo  {journal} {Phys. Rev. B}\ }\textbf {\bibinfo {volume}
  {91}},\ \bibinfo {pages} {115425} (\bibinfo {year} {2015})}\BibitemShut
  {NoStop}%
\bibitem [{\citenamefont {Ryabov}\ and\ \citenamefont
  {Holubec}(2016)}]{Ryabov2016}%
  \BibitemOpen
  \bibfield  {author} {\bibinfo {author} {\bibfnamefont {A.}~\bibnamefont
  {Ryabov}}\ and\ \bibinfo {author} {\bibfnamefont {V.}~\bibnamefont
  {Holubec}},\ }\href {\doibase 10.1103/PhysRevE.93.050101} {\bibfield
  {journal} {\bibinfo  {journal} {Phys. Rev. E}\ }\textbf {\bibinfo {volume}
  {93}},\ \bibinfo {pages} {050101} (\bibinfo {year} {2016})}\BibitemShut
  {NoStop}%
\bibitem [{\citenamefont {Zhang}\ \emph {et~al.}(2017)\citenamefont {Zhang},
  \citenamefont {Li}, \citenamefont {Tang}, \citenamefont {Yang},\ and\
  \citenamefont {Bai}}]{ThermoelectricZhangPRE2017}%
  \BibitemOpen
  \bibfield  {author} {\bibinfo {author} {\bibfnamefont {R.}~\bibnamefont
  {Zhang}}, \bibinfo {author} {\bibfnamefont {Q.-W.}\ \bibnamefont {Li}},
  \bibinfo {author} {\bibfnamefont {F.~R.}\ \bibnamefont {Tang}}, \bibinfo
  {author} {\bibfnamefont {X.~Q.}\ \bibnamefont {Yang}}, \ and\ \bibinfo
  {author} {\bibfnamefont {L.}~\bibnamefont {Bai}},\ }\href {\doibase
  10.1103/PhysRevE.96.022133} {\bibfield  {journal} {\bibinfo  {journal} {Phys.
  Rev. E}\ }\textbf {\bibinfo {volume} {96}},\ \bibinfo {pages} {022133}
  (\bibinfo {year} {2017})}\BibitemShut {NoStop}%
\bibitem [{\citenamefont {Holubec}\ and\ \citenamefont
  {Ryabov}(2016{\natexlab{a}})}]{Holubec2016}%
  \BibitemOpen
  \bibfield  {author} {\bibinfo {author} {\bibfnamefont {V.}~\bibnamefont
  {Holubec}}\ and\ \bibinfo {author} {\bibfnamefont {A.}~\bibnamefont
  {Ryabov}},\ }\href {http://stacks.iop.org/1742-5468/2016/i=7/a=073204}
  {\bibfield  {journal} {\bibinfo  {journal} {J. Stat. Mech: Theory Exp.}\
  }\textbf {\bibinfo {volume} {2016}},\ \bibinfo {pages} {073204} (\bibinfo
  {year} {2016}{\natexlab{a}})}\BibitemShut {NoStop}%
\bibitem [{\citenamefont {Long}\ and\ \citenamefont {Liu}(2016)}]{Long2016}%
  \BibitemOpen
  \bibfield  {author} {\bibinfo {author} {\bibfnamefont {R.}~\bibnamefont
  {Long}}\ and\ \bibinfo {author} {\bibfnamefont {W.}~\bibnamefont {Liu}},\
  }\href {\doibase 10.1103/PhysRevE.94.052114} {\bibfield  {journal} {\bibinfo
  {journal} {Phys. Rev. E}\ }\textbf {\bibinfo {volume} {94}},\ \bibinfo
  {pages} {052114} (\bibinfo {year} {2016})}\BibitemShut {NoStop}%
\bibitem [{\citenamefont {Dechant}\ \emph {et~al.}(2016)\citenamefont
  {Dechant}, \citenamefont {Kiesel},\ and\ \citenamefont {Lutz}}]{Dechant2016}%
  \BibitemOpen
  \bibfield  {author} {\bibinfo {author} {\bibfnamefont {A.}~\bibnamefont
  {Dechant}}, \bibinfo {author} {\bibfnamefont {N.}~\bibnamefont {Kiesel}}, \
  and\ \bibinfo {author} {\bibfnamefont {E.}~\bibnamefont {Lutz}},\ }\href@noop
  {} {\bibfield  {journal} {\bibinfo  {journal} {arXiv:1602.00392}\ } (\bibinfo
  {year} {2016})}\BibitemShut {NoStop}%
\bibitem [{\citenamefont {Martinez}\ \emph {et~al.}(2017)\citenamefont
  {Martinez}, \citenamefont {Roldan}, \citenamefont {Dinis},\ and\
  \citenamefont {Rica}}]{Martinez2017}%
  \BibitemOpen
  \bibfield  {author} {\bibinfo {author} {\bibfnamefont {I.~A.}\ \bibnamefont
  {Martinez}}, \bibinfo {author} {\bibfnamefont {E.}~\bibnamefont {Roldan}},
  \bibinfo {author} {\bibfnamefont {L.}~\bibnamefont {Dinis}}, \ and\ \bibinfo
  {author} {\bibfnamefont {R.~A.}\ \bibnamefont {Rica}},\ }\href {\doibase
  10.1039/C6SM00923A} {\bibfield  {journal} {\bibinfo  {journal} {Soft Matter}\
  }\textbf {\bibinfo {volume} {13}},\ \bibinfo {pages} {22} (\bibinfo {year}
  {2017})}\BibitemShut {NoStop}%
\bibitem [{\citenamefont {Blickle}\ and\ \citenamefont
  {Bechinger}(2012)}]{Blickle2012}%
  \BibitemOpen
  \bibfield  {author} {\bibinfo {author} {\bibfnamefont {V.}~\bibnamefont
  {Blickle}}\ and\ \bibinfo {author} {\bibfnamefont {C.}~\bibnamefont
  {Bechinger}},\ }\href
  {http://www.nature.com/nphys/journal/v8/n2/abs/nphys2163.html} {\bibfield
  {journal} {\bibinfo  {journal} {Nat. Phys.}\ }\textbf {\bibinfo {volume}
  {8}},\ \bibinfo {pages} {143} (\bibinfo {year} {2012})}\BibitemShut {NoStop}%
\bibitem [{\citenamefont {Mart{\'\i}nez}\ \emph {et~al.}(2016)\citenamefont
  {Mart{\'\i}nez}, \citenamefont {Rold{\'a}n}, \citenamefont {Dinis},
  \citenamefont {Petrov}, \citenamefont {Parrondo},\ and\ \citenamefont
  {Rica}}]{Martinez2016}%
  \BibitemOpen
  \bibfield  {author} {\bibinfo {author} {\bibfnamefont {I.~A.}\ \bibnamefont
  {Mart{\'\i}nez}}, \bibinfo {author} {\bibfnamefont {{\'E}.}~\bibnamefont
  {Rold{\'a}n}}, \bibinfo {author} {\bibfnamefont {L.}~\bibnamefont {Dinis}},
  \bibinfo {author} {\bibfnamefont {D.}~\bibnamefont {Petrov}}, \bibinfo
  {author} {\bibfnamefont {J.~M.}\ \bibnamefont {Parrondo}}, \ and\ \bibinfo
  {author} {\bibfnamefont {R.~A.}\ \bibnamefont {Rica}},\ }\href
  {http://www.nature.com/nphys/journal/v12/n1/abs/nphys3518.html} {\bibfield
  {journal} {\bibinfo  {journal} {Nat. Phys.}\ }\textbf {\bibinfo {volume}
  {12}},\ \bibinfo {pages} {67} (\bibinfo {year} {2016})}\BibitemShut {NoStop}%
\bibitem [{\citenamefont {Gavrilov}\ and\ \citenamefont
  {Bechhoefer}(2016{\natexlab{a}})}]{Gavrilov2016}%
  \BibitemOpen
  \bibfield  {author} {\bibinfo {author} {\bibfnamefont {M.}~\bibnamefont
  {Gavrilov}}\ and\ \bibinfo {author} {\bibfnamefont {J.}~\bibnamefont
  {Bechhoefer}},\ }\href {\doibase 10.1103/PhysRevLett.117.200601} {\bibfield
  {journal} {\bibinfo  {journal} {Phys. Rev. Lett.}\ }\textbf {\bibinfo
  {volume} {117}},\ \bibinfo {pages} {200601} (\bibinfo {year}
  {2016}{\natexlab{a}})}\BibitemShut {NoStop}%
\bibitem [{\citenamefont {Cohen}\ and\ \citenamefont
  {Moerner}(2005)}]{Cohen2005}%
  \BibitemOpen
  \bibfield  {author} {\bibinfo {author} {\bibfnamefont {A.~E.}\ \bibnamefont
  {Cohen}}\ and\ \bibinfo {author} {\bibfnamefont {W.~E.}\ \bibnamefont
  {Moerner}},\ }\href {\doibase 10.1063/1.1872220} {\bibfield  {journal}
  {\bibinfo  {journal} {Appl. Phys. Lett.}\ }\textbf {\bibinfo {volume} {86}},\
  \bibinfo {pages} {093109} (\bibinfo {year} {2005})}\BibitemShut {NoStop}%
\bibitem [{\citenamefont {Jun}\ and\ \citenamefont
  {Bechhoefer}(2012)}]{Jun2012}%
  \BibitemOpen
  \bibfield  {author} {\bibinfo {author} {\bibfnamefont {Y.}~\bibnamefont
  {Jun}}\ and\ \bibinfo {author} {\bibfnamefont {J.}~\bibnamefont
  {Bechhoefer}},\ }\href {\doibase 10.1103/PhysRevE.86.061106} {\bibfield
  {journal} {\bibinfo  {journal} {Phys. Rev. E}\ }\textbf {\bibinfo {volume}
  {86}},\ \bibinfo {pages} {061106} (\bibinfo {year} {2012})}\BibitemShut
  {NoStop}%
\bibitem [{\citenamefont {Cohen}(2005)}]{Cohen2005a}%
  \BibitemOpen
  \bibfield  {author} {\bibinfo {author} {\bibfnamefont {A.~E.}\ \bibnamefont
  {Cohen}},\ }\href {\doibase 10.1103/PhysRevLett.94.118102} {\bibfield
  {journal} {\bibinfo  {journal} {Phys. Rev. Lett.}\ }\textbf {\bibinfo
  {volume} {94}},\ \bibinfo {pages} {118102} (\bibinfo {year}
  {2005})}\BibitemShut {NoStop}%
\bibitem [{\citenamefont {Gavrilov}\ and\ \citenamefont
  {Bechhoefer}(2016{\natexlab{b}})}]{Gavrilov2016a}%
  \BibitemOpen
  \bibfield  {author} {\bibinfo {author} {\bibfnamefont {M.}~\bibnamefont
  {Gavrilov}}\ and\ \bibinfo {author} {\bibfnamefont {J.}~\bibnamefont
  {Bechhoefer}},\ }\href {\doibase 10.1209/0295-5075/114/50002} {\bibfield
  {journal} {\bibinfo  {journal} {EPL}\ }\textbf {\bibinfo {volume} {114}},\
  \bibinfo {pages} {50002} (\bibinfo {year} {2016}{\natexlab{b}})}\BibitemShut
  {NoStop}%
\bibitem [{\citenamefont {Krishnamurthy}\ \emph {et~al.}(2016)\citenamefont
  {Krishnamurthy}, \citenamefont {Ghosh}, \citenamefont {Chatterji},
  \citenamefont {Ganapathy},\ and\ \citenamefont {Sood}}]{Krishnamurthy2016}%
  \BibitemOpen
  \bibfield  {author} {\bibinfo {author} {\bibfnamefont {S.}~\bibnamefont
  {Krishnamurthy}}, \bibinfo {author} {\bibfnamefont {S.}~\bibnamefont
  {Ghosh}}, \bibinfo {author} {\bibfnamefont {D.}~\bibnamefont {Chatterji}},
  \bibinfo {author} {\bibfnamefont {R.}~\bibnamefont {Ganapathy}}, \ and\
  \bibinfo {author} {\bibfnamefont {A.~K.}\ \bibnamefont {Sood}},\ }\href
  {http://dx.doi.org/10.1038/nphys3870} {\bibfield  {journal} {\bibinfo
  {journal} {Nat. Phys.}\ }\textbf {\bibinfo {volume} {12}},\ \bibinfo {pages}
  {1134} (\bibinfo {year} {2016})}\BibitemShut {NoStop}%
\bibitem [{\citenamefont {Holubec}\ and\ \citenamefont
  {Ryabov}(2015)}]{HolubecRyabov2015}%
  \BibitemOpen
  \bibfield  {author} {\bibinfo {author} {\bibfnamefont {V.}~\bibnamefont
  {Holubec}}\ and\ \bibinfo {author} {\bibfnamefont {A.}~\bibnamefont
  {Ryabov}},\ }\href {\doibase 10.1103/PhysRevE.92.052125} {\bibfield
  {journal} {\bibinfo  {journal} {Phys. Rev. E}\ }\textbf {\bibinfo {volume}
  {92}},\ \bibinfo {pages} {052125} (\bibinfo {year} {2015})}\BibitemShut
  {NoStop}%
\bibitem [{\citenamefont {Holubec}\ and\ \citenamefont
  {Ryabov}(2016{\natexlab{b}})}]{Holubec2016a}%
  \BibitemOpen
  \bibfield  {author} {\bibinfo {author} {\bibfnamefont {V.}~\bibnamefont
  {Holubec}}\ and\ \bibinfo {author} {\bibfnamefont {A.}~\bibnamefont
  {Ryabov}},\ }\href {\doibase 10.1103/PhysRevE.93.059904} {\bibfield
  {journal} {\bibinfo  {journal} {Phys. Rev. E}\ }\textbf {\bibinfo {volume}
  {93}},\ \bibinfo {pages} {059904} (\bibinfo {year}
  {2016}{\natexlab{b}})}\BibitemShut {NoStop}%
\bibitem [{\citenamefont {Gordon}\ and\ \citenamefont
  {Huleihil}(1992)}]{Gordon1992}%
  \BibitemOpen
  \bibfield  {author} {\bibinfo {author} {\bibfnamefont {J.~M.}\ \bibnamefont
  {Gordon}}\ and\ \bibinfo {author} {\bibfnamefont {M.}~\bibnamefont
  {Huleihil}},\ }\href {\doibase 10.1063/1.351755} {\bibfield  {journal}
  {\bibinfo  {journal} {J. Appl. Phys.}\ }\textbf {\bibinfo {volume} {72}},\
  \bibinfo {pages} {829} (\bibinfo {year} {1992})}\BibitemShut {NoStop}%
\bibitem [{\citenamefont {Baldasaro}\ \emph {et~al.}(2001)\citenamefont
  {Baldasaro}, \citenamefont {Raynolds}, \citenamefont {Charache},
  \citenamefont {DePoy}, \citenamefont {Ballinger}, \citenamefont {Donovan},\
  and\ \citenamefont {Borrego}}]{Baldasaro2001}%
  \BibitemOpen
  \bibfield  {author} {\bibinfo {author} {\bibfnamefont {P.~F.}\ \bibnamefont
  {Baldasaro}}, \bibinfo {author} {\bibfnamefont {J.~E.}\ \bibnamefont
  {Raynolds}}, \bibinfo {author} {\bibfnamefont {G.~W.}\ \bibnamefont
  {Charache}}, \bibinfo {author} {\bibfnamefont {D.~M.}\ \bibnamefont {DePoy}},
  \bibinfo {author} {\bibfnamefont {C.~T.}\ \bibnamefont {Ballinger}}, \bibinfo
  {author} {\bibfnamefont {T.}~\bibnamefont {Donovan}}, \ and\ \bibinfo
  {author} {\bibfnamefont {J.~M.}\ \bibnamefont {Borrego}},\ }\href {\doibase
  10.1063/1.1344580} {\bibfield  {journal} {\bibinfo  {journal} {J. Appl.
  Phys.}\ }\textbf {\bibinfo {volume} {89}},\ \bibinfo {pages} {3319} (\bibinfo
  {year} {2001})}\BibitemShut {NoStop}%
\bibitem [{\citenamefont {Smith}(2004)}]{Smith2004}%
  \BibitemOpen
  \bibfield  {author} {\bibinfo {author} {\bibfnamefont {T.~C.}\ \bibnamefont
  {Smith}},\ }in\ \href@noop {} {\emph {\bibinfo {booktitle} {Proceedings of
  the 2nd International Energy Conversion Engineering Conference, Providence
  (RI)}}}\ (\bibinfo {year} {2004})\ pp.\ \bibinfo {pages} {1--15}\BibitemShut
  {NoStop}%
\bibitem [{\citenamefont {{del Campo}}\ \emph {et~al.}(2014)\citenamefont {{del
  Campo}}, \citenamefont {Goold},\ and\ \citenamefont
  {Paternostro}}]{DelCampo2014}%
  \BibitemOpen
  \bibfield  {author} {\bibinfo {author} {\bibfnamefont {A.}~\bibnamefont {{del
  Campo}}}, \bibinfo {author} {\bibfnamefont {J.}~\bibnamefont {Goold}}, \ and\
  \bibinfo {author} {\bibfnamefont {M.}~\bibnamefont {Paternostro}},\ }\href
  {http://dx.doi.org/10.1038/srep06208} {\bibfield  {journal} {\bibinfo
  {journal} {Sci. Rep.}\ }\textbf {\bibinfo {volume} {4}},\ \bibinfo {pages}
  {6208} (\bibinfo {year} {2014})}\BibitemShut {NoStop}%
\bibitem [{\citenamefont {Long}\ and\ \citenamefont {Liu}(2015)}]{Long2015}%
  \BibitemOpen
  \bibfield  {author} {\bibinfo {author} {\bibfnamefont {R.}~\bibnamefont
  {Long}}\ and\ \bibinfo {author} {\bibfnamefont {W.}~\bibnamefont {Liu}},\
  }\href {\doibase 10.1103/PhysRevE.91.042127} {\bibfield  {journal} {\bibinfo
  {journal} {Phys. Rev. E}\ }\textbf {\bibinfo {volume} {91}},\ \bibinfo
  {pages} {042127} (\bibinfo {year} {2015})}\BibitemShut {NoStop}%
\bibitem [{\citenamefont {Haseli}(2013)}]{Haseli2013}%
  \BibitemOpen
  \bibfield  {author} {\bibinfo {author} {\bibfnamefont {Y.}~\bibnamefont
  {Haseli}},\ }\href {\doibase
  http://dx.doi.org/10.1016/j.enconman.2012.12.033} {\bibfield  {journal}
  {\bibinfo  {journal} {Energ. Convers. Manage.}\ }\textbf {\bibinfo {volume}
  {68}},\ \bibinfo {pages} {133 } (\bibinfo {year} {2013})}\BibitemShut
  {NoStop}%
\bibitem [{\citenamefont {Gonzalez-Ayala}\ \emph
  {et~al.}(2017{\natexlab{a}})\citenamefont {Gonzalez-Ayala}, \citenamefont
  {Calvo~Hern{\' a}ndez},\ and\ \citenamefont {Roco}}]{Gonzalez-AyalaPRE2017}%
  \BibitemOpen
  \bibfield  {author} {\bibinfo {author} {\bibfnamefont {J.}~\bibnamefont
  {Gonzalez-Ayala}}, \bibinfo {author} {\bibfnamefont {A.}~\bibnamefont
  {Calvo~Hern{\' a}ndez}}, \ and\ \bibinfo {author} {\bibfnamefont {J.~M.~M.}\
  \bibnamefont {Roco}},\ }\href {\doibase 10.1103/PhysRevE.95.022131}
  {\bibfield  {journal} {\bibinfo  {journal} {Phys. Rev. E}\ }\textbf {\bibinfo
  {volume} {95}},\ \bibinfo {pages} {022131} (\bibinfo {year}
  {2017}{\natexlab{a}})}\BibitemShut {NoStop}%
\bibitem [{\citenamefont {Wang}(2016)}]{WangEntropy2017}%
  \BibitemOpen
  \bibfield  {author} {\bibinfo {author} {\bibfnamefont {Y.}~\bibnamefont
  {Wang}},\ }\href {\doibase 10.3390/e18050161} {\bibfield  {journal} {\bibinfo
   {journal} {Entropy}\ }\textbf {\bibinfo {volume} {18}},\ \bibinfo {pages}
  {161} (\bibinfo {year} {2016})}\BibitemShut {NoStop}%
\bibitem [{\citenamefont {Benenti}\ \emph {et~al.}(2017)\citenamefont
  {Benenti}, \citenamefont {Casati}, \citenamefont {Saito},\ and\ \citenamefont
  {Whitney}}]{Benenti2017}%
  \BibitemOpen
  \bibfield  {author} {\bibinfo {author} {\bibfnamefont {G.}~\bibnamefont
  {Benenti}}, \bibinfo {author} {\bibfnamefont {G.}~\bibnamefont {Casati}},
  \bibinfo {author} {\bibfnamefont {K.}~\bibnamefont {Saito}}, \ and\ \bibinfo
  {author} {\bibfnamefont {R.~S.}\ \bibnamefont {Whitney}},\ }\href {\doibase
  https://doi.org/10.1016/j.physrep.2017.05.008} {\bibfield  {journal}
  {\bibinfo  {journal} {Phys. Rep.}\ }\textbf {\bibinfo {volume} {694}},\
  \bibinfo {pages} {1 } (\bibinfo {year} {2017})}\BibitemShut {NoStop}%
\bibitem [{\citenamefont {Raz}\ \emph {et~al.}(2016)\citenamefont {Raz},
  \citenamefont {Suba{\c{s}}i},\ and\ \citenamefont {Jarzynski}}]{Raz2016}%
  \BibitemOpen
  \bibfield  {author} {\bibinfo {author} {\bibfnamefont {O.}~\bibnamefont
  {Raz}}, \bibinfo {author} {\bibfnamefont {Y.}~\bibnamefont {Suba{\c{s}}i}}, \
  and\ \bibinfo {author} {\bibfnamefont {C.}~\bibnamefont {Jarzynski}},\ }\href
  {\doibase 10.1103/PhysRevX.6.021022} {\bibfield  {journal} {\bibinfo
  {journal} {Phys. Rev. X}\ }\textbf {\bibinfo {volume} {6}},\ \bibinfo {pages}
  {021022} (\bibinfo {year} {2016})}\BibitemShut {NoStop}%
\bibitem [{\citenamefont {Brandner}\ and\ \citenamefont
  {Seifert}(2015)}]{Brandner2015a}%
  \BibitemOpen
  \bibfield  {author} {\bibinfo {author} {\bibfnamefont {K.}~\bibnamefont
  {Brandner}}\ and\ \bibinfo {author} {\bibfnamefont {U.}~\bibnamefont
  {Seifert}},\ }\href {\doibase 10.1103/PhysRevE.91.012121} {\bibfield
  {journal} {\bibinfo  {journal} {Phys. Rev. E}\ }\textbf {\bibinfo {volume}
  {91}},\ \bibinfo {pages} {012121} (\bibinfo {year} {2015})}\BibitemShut
  {NoStop}%
\bibitem [{\citenamefont {Van~den Broeck}(2005)}]{BroeckPRL2005}%
  \BibitemOpen
  \bibfield  {author} {\bibinfo {author} {\bibfnamefont {C.}~\bibnamefont
  {Van~den Broeck}},\ }\href {\doibase 10.1103/PhysRevLett.95.190602}
  {\bibfield  {journal} {\bibinfo  {journal} {Phys. Rev. Lett.}\ }\textbf
  {\bibinfo {volume} {95}},\ \bibinfo {pages} {190602} (\bibinfo {year}
  {2005})}\BibitemShut {NoStop}%
\bibitem [{\citenamefont {Polettini}\ \emph
  {et~al.}(2015{\natexlab{b}})\citenamefont {Polettini}, \citenamefont
  {Verley},\ and\ \citenamefont {Esposito}}]{Polettini2015a}%
  \BibitemOpen
  \bibfield  {author} {\bibinfo {author} {\bibfnamefont {M.}~\bibnamefont
  {Polettini}}, \bibinfo {author} {\bibfnamefont {G.}~\bibnamefont {Verley}}, \
  and\ \bibinfo {author} {\bibfnamefont {M.}~\bibnamefont {Esposito}},\ }\href
  {\doibase 10.1103/PhysRevLett.114.050601} {\bibfield  {journal} {\bibinfo
  {journal} {Phys. Rev. Lett.}\ }\textbf {\bibinfo {volume} {114}},\ \bibinfo
  {pages} {050601} (\bibinfo {year} {2015}{\natexlab{b}})}\BibitemShut
  {NoStop}%
\bibitem [{\citenamefont {Binder}\ \emph {et~al.}(2007)\citenamefont {Binder},
  \citenamefont {Das}, \citenamefont {Fisher}, \citenamefont {Horbach},\ and\
  \citenamefont {Sengers}}]{Binder2007}%
  \BibitemOpen
  \bibfield  {author} {\bibinfo {author} {\bibfnamefont {K.}~\bibnamefont
  {Binder}}, \bibinfo {author} {\bibfnamefont {S.~K.}\ \bibnamefont {Das}},
  \bibinfo {author} {\bibfnamefont {M.~E.}\ \bibnamefont {Fisher}}, \bibinfo
  {author} {\bibfnamefont {J.}~\bibnamefont {Horbach}}, \ and\ \bibinfo
  {author} {\bibfnamefont {J.~V.}\ \bibnamefont {Sengers}},\ }in\ \href
  {http://elib.dlr.de/52154/} {\emph {\bibinfo {booktitle} {Diffusion
  Fundamentals II}}},\ \bibinfo {editor} {edited by\ \bibinfo {editor}
  {\bibfnamefont {S.}~\bibnamefont {Brandani}}, \bibinfo {editor}
  {\bibfnamefont {C.}~\bibnamefont {Chmelik}}, \bibinfo {editor} {\bibfnamefont
  {J.}~\bibnamefont {K{\"a}rger}}, \ and\ \bibinfo {editor} {\bibfnamefont
  {R.}~\bibnamefont {Volpe}}}\ (\bibinfo  {publisher} {Leipziger
  Universit{\"a}tsverlag},\ \bibinfo {year} {2007})\ pp.\ \bibinfo {pages}
  {120--131}\BibitemShut {NoStop}%
\bibitem [{\citenamefont {Izumida}\ and\ \citenamefont
  {Okuda}(2012)}]{Izumida2012}%
  \BibitemOpen
  \bibfield  {author} {\bibinfo {author} {\bibfnamefont {Y.}~\bibnamefont
  {Izumida}}\ and\ \bibinfo {author} {\bibfnamefont {K.}~\bibnamefont
  {Okuda}},\ }\href {http://stacks.iop.org/0295-5075/97/i=1/a=10004} {\bibfield
   {journal} {\bibinfo  {journal} {EPL}\ }\textbf {\bibinfo {volume} {97}},\
  \bibinfo {pages} {10004} (\bibinfo {year} {2012})}\BibitemShut {NoStop}%
\bibitem [{\citenamefont {Esposito}\ \emph {et~al.}(2010)\citenamefont
  {Esposito}, \citenamefont {Kawai}, \citenamefont {Lindenberg},\ and\
  \citenamefont {Van~den Broeck}}]{Esposito2010b}%
  \BibitemOpen
  \bibfield  {author} {\bibinfo {author} {\bibfnamefont {M.}~\bibnamefont
  {Esposito}}, \bibinfo {author} {\bibfnamefont {R.}~\bibnamefont {Kawai}},
  \bibinfo {author} {\bibfnamefont {K.}~\bibnamefont {Lindenberg}}, \ and\
  \bibinfo {author} {\bibfnamefont {C.}~\bibnamefont {Van~den Broeck}},\ }\href
  {\doibase 10.1103/PhysRevLett.105.150603} {\bibfield  {journal} {\bibinfo
  {journal} {Phys. Rev. Lett.}\ }\textbf {\bibinfo {volume} {105}},\ \bibinfo
  {pages} {150603} (\bibinfo {year} {2010})}\BibitemShut {NoStop}%
\bibitem [{\citenamefont {Schmiedl}\ and\ \citenamefont
  {Seifert}(2008)}]{Schmiedl2008}%
  \BibitemOpen
  \bibfield  {author} {\bibinfo {author} {\bibfnamefont {T.}~\bibnamefont
  {Schmiedl}}\ and\ \bibinfo {author} {\bibfnamefont {U.}~\bibnamefont
  {Seifert}},\ }\href {http://stacks.iop.org/0295-5075/81/i=2/a=20003}
  {\bibfield  {journal} {\bibinfo  {journal} {EPL}\ }\textbf {\bibinfo {volume}
  {81}},\ \bibinfo {pages} {20003} (\bibinfo {year} {2008})}\BibitemShut
  {NoStop}%
\bibitem [{\citenamefont {Holubec}(2014)}]{Holubec2014}%
  \BibitemOpen
  \bibfield  {author} {\bibinfo {author} {\bibfnamefont {V.}~\bibnamefont
  {Holubec}},\ }\href {http://stacks.iop.org/1742-5468/2014/i=5/a=P05022}
  {\bibfield  {journal} {\bibinfo  {journal} {J. Stat. Mech: Theory Exp.}\
  }\textbf {\bibinfo {volume} {2014}},\ \bibinfo {pages} {P05022} (\bibinfo
  {year} {2014})}\BibitemShut {NoStop}%
\bibitem [{\citenamefont {Gonzalez-Ayala}\ \emph
  {et~al.}(2017{\natexlab{b}})\citenamefont {Gonzalez-Ayala}, \citenamefont
  {Roco}, \citenamefont {Medina},\ and\ \citenamefont {Calvo~Hern{\'
  a}ndez}}]{Gonzalez-AyalaEntropy2017}%
  \BibitemOpen
  \bibfield  {author} {\bibinfo {author} {\bibfnamefont {J.}~\bibnamefont
  {Gonzalez-Ayala}}, \bibinfo {author} {\bibfnamefont {J.~M.~M.}\ \bibnamefont
  {Roco}}, \bibinfo {author} {\bibfnamefont {A.}~\bibnamefont {Medina}}, \ and\
  \bibinfo {author} {\bibfnamefont {A.}~\bibnamefont {Calvo~Hern{\' a}ndez}},\
  }\href {\doibase 10.3390/e19040182} {\bibfield  {journal} {\bibinfo
  {journal} {Entropy}\ }\textbf {\bibinfo {volume} {19}},\ \bibinfo {pages}
  {182} (\bibinfo {year} {2017}{\natexlab{b}})}\BibitemShut {NoStop}%
\bibitem [{\citenamefont {Johal}(2017)}]{Johal2017PRE}%
  \BibitemOpen
  \bibfield  {author} {\bibinfo {author} {\bibfnamefont {R.~S.}\ \bibnamefont
  {Johal}},\ }\href {\doibase 10.1103/PhysRevE.96.012151} {\bibfield  {journal}
  {\bibinfo  {journal} {Phys. Rev. E}\ }\textbf {\bibinfo {volume} {96}},\
  \bibinfo {pages} {012151} (\bibinfo {year} {2017})}\BibitemShut {NoStop}%
\bibitem [{\citenamefont {Zulkowski}\ and\ \citenamefont
  {DeWeese}(2015{\natexlab{a}})}]{Zulkowski2015a}%
  \BibitemOpen
  \bibfield  {author} {\bibinfo {author} {\bibfnamefont {P.~R.}\ \bibnamefont
  {Zulkowski}}\ and\ \bibinfo {author} {\bibfnamefont {M.~R.}\ \bibnamefont
  {DeWeese}},\ }\href {\doibase 10.1103/PhysRevE.92.032117} {\bibfield
  {journal} {\bibinfo  {journal} {Phys. Rev. E}\ }\textbf {\bibinfo {volume}
  {92}},\ \bibinfo {pages} {032117} (\bibinfo {year}
  {2015}{\natexlab{a}})}\BibitemShut {NoStop}%
\bibitem [{\citenamefont {Zulkowski}\ and\ \citenamefont
  {DeWeese}(2015{\natexlab{b}})}]{Zulkowski2015}%
  \BibitemOpen
  \bibfield  {author} {\bibinfo {author} {\bibfnamefont {P.~R.}\ \bibnamefont
  {Zulkowski}}\ and\ \bibinfo {author} {\bibfnamefont {M.~R.}\ \bibnamefont
  {DeWeese}},\ }\href {\doibase 10.1103/PhysRevE.92.032113} {\bibfield
  {journal} {\bibinfo  {journal} {Phys. Rev. E}\ }\textbf {\bibinfo {volume}
  {92}},\ \bibinfo {pages} {032113} (\bibinfo {year}
  {2015}{\natexlab{b}})}\BibitemShut {NoStop}%
\bibitem [{\citenamefont {Sato}\ \emph {et~al.}(2002)\citenamefont {Sato},
  \citenamefont {Sekimoto}, \citenamefont {Hondou},\ and\ \citenamefont
  {Takagi}}]{Sato2002}%
  \BibitemOpen
  \bibfield  {author} {\bibinfo {author} {\bibfnamefont {K.}~\bibnamefont
  {Sato}}, \bibinfo {author} {\bibfnamefont {K.}~\bibnamefont {Sekimoto}},
  \bibinfo {author} {\bibfnamefont {T.}~\bibnamefont {Hondou}}, \ and\ \bibinfo
  {author} {\bibfnamefont {F.}~\bibnamefont {Takagi}},\ }\href {\doibase
  10.1103/PhysRevE.66.016119} {\bibfield  {journal} {\bibinfo  {journal} {Phys.
  Rev. E}\ }\textbf {\bibinfo {volume} {66}},\ \bibinfo {pages} {016119}
  (\bibinfo {year} {2002})}\BibitemShut {NoStop}%
\bibitem [{\citenamefont {Curzon}\ and\ \citenamefont
  {Ahlborn}(1975)}]{Curzon1975}%
  \BibitemOpen
  \bibfield  {author} {\bibinfo {author} {\bibfnamefont {F.~L.}\ \bibnamefont
  {Curzon}}\ and\ \bibinfo {author} {\bibfnamefont {B.}~\bibnamefont
  {Ahlborn}},\ }\href {\doibase 10.1119/1.10023} {\bibfield  {journal}
  {\bibinfo  {journal} {Am. J. Phys.}\ }\textbf {\bibinfo {volume} {43}},\
  \bibinfo {pages} {22} (\bibinfo {year} {1975})}\BibitemShut {NoStop}%
\bibitem [{\citenamefont {Chambadal}(1957)}]{Chambadal1957}%
  \BibitemOpen
  \bibfield  {author} {\bibinfo {author} {\bibfnamefont {P.}~\bibnamefont
  {Chambadal}},\ }\href@noop {} {\emph {\bibinfo {title} {Les centrales
  nucl{\'e}aires}}},\ Vol.\ \bibinfo {volume} {321}\ (\bibinfo  {publisher}
  {Colin},\ \bibinfo {year} {1957})\BibitemShut {NoStop}%
\bibitem [{\citenamefont {Novikov}(1958)}]{Novikov1958}%
  \BibitemOpen
  \bibfield  {author} {\bibinfo {author} {\bibfnamefont {I.~I.}\ \bibnamefont
  {Novikov}},\ }\href@noop {} {\bibfield  {journal} {\bibinfo  {journal} {J.
  Nucl. Energy II}\ }\textbf {\bibinfo {volume} {7}},\ \bibinfo {pages} {125}
  (\bibinfo {year} {1958})}\BibitemShut {NoStop}%
\bibitem [{\citenamefont {Yvon}(1955)}]{Yvon1955}%
  \BibitemOpen
  \bibfield  {author} {\bibinfo {author} {\bibfnamefont {J.}~\bibnamefont
  {Yvon}},\ }in\ \href@noop {} {\emph {\bibinfo {booktitle} {Proceedings of the
  International Conference on Peaceful Uses of Atomic Energy}}}\ (\bibinfo
  {address} {Geneva},\ \bibinfo {year} {1955})\ p.\ \bibinfo {pages}
  {387}\BibitemShut {NoStop}%
\bibitem [{\citenamefont {Risken}\ and\ \citenamefont
  {Frank}(1996)}]{Risken1996}%
  \BibitemOpen
  \bibfield  {author} {\bibinfo {author} {\bibfnamefont {H.}~\bibnamefont
  {Risken}}\ and\ \bibinfo {author} {\bibfnamefont {T.}~\bibnamefont {Frank}},\
  }\href {https://books.google.de/books?id=MG2V9vTgSgEC} {\emph {\bibinfo
  {title} {The Fokker-Planck Equation: Methods of Solution and
  Applications}}},\ Springer Series in Synergetics\ (\bibinfo  {publisher}
  {Springer Berlin Heidelberg},\ \bibinfo {year} {1996})\BibitemShut {NoStop}%
\bibitem [{\citenamefont {Pigolotti}\ \emph {et~al.}(2017)\citenamefont
  {Pigolotti}, \citenamefont {Neri}, \citenamefont {Rold\'an},\ and\
  \citenamefont {J\"ulicher}}]{Pigolotti2017}%
  \BibitemOpen
  \bibfield  {author} {\bibinfo {author} {\bibfnamefont {S.}~\bibnamefont
  {Pigolotti}}, \bibinfo {author} {\bibfnamefont {I.}~\bibnamefont {Neri}},
  \bibinfo {author} {\bibfnamefont {E.}~\bibnamefont {Rold\'an}}, \ and\
  \bibinfo {author} {\bibfnamefont {F.}~\bibnamefont {J\"ulicher}},\ }\href
  {\doibase 10.1103/PhysRevLett.119.140604} {\bibfield  {journal} {\bibinfo
  {journal} {Phys. Rev. Lett.}\ }\textbf {\bibinfo {volume} {119}},\ \bibinfo
  {pages} {140604} (\bibinfo {year} {2017})}\BibitemShut {NoStop}%
\bibitem [{Note1()}]{Note1}%
  \BibitemOpen
  \bibinfo {note} {Although the presented thermodynamic analysis of the
  Brownian HE is standard, it neglects heat currents connected with momentum
  degrees of freedom \cite {Schmiedl2008,Martinez2015, Arold2017}. These heat
  currents inevitably lower the efficiency. In the over-damped limit, the
  momentum is assumed to be in equilibrium and the effect of the corresponding
  heat flow on the efficiency can be canceled out by introducing a regenerator.
  If this is not possible, one should use different optimal protocols than
  those derived in the over-damped limit, especially with the aim to facilitate
  the additional heat flux into the momentum space. Such optimization procedure
  cannot be performed analytically and is out of the scope of the present
  paper.}\BibitemShut {Stop}%
\bibitem [{\citenamefont {Seifert}(2011)}]{Seifert2011}%
  \BibitemOpen
  \bibfield  {author} {\bibinfo {author} {\bibfnamefont {U.}~\bibnamefont
  {Seifert}},\ }\href {\doibase 10.1103/PhysRevLett.106.020601} {\bibfield
  {journal} {\bibinfo  {journal} {Phys. Rev. Lett.}\ }\textbf {\bibinfo
  {volume} {106}},\ \bibinfo {pages} {020601} (\bibinfo {year}
  {2011})}\BibitemShut {NoStop}%
\bibitem [{\citenamefont {Mart\'{\i}nez}\ \emph {et~al.}(2015)\citenamefont
  {Mart\'{\i}nez}, \citenamefont {Rold\'an}, \citenamefont {Dinis},
  \citenamefont {Petrov},\ and\ \citenamefont {Rica}}]{Martinez2015}%
  \BibitemOpen
  \bibfield  {author} {\bibinfo {author} {\bibfnamefont {I.~A.}\ \bibnamefont
  {Mart\'{\i}nez}}, \bibinfo {author} {\bibfnamefont {E.}~\bibnamefont
  {Rold\'an}}, \bibinfo {author} {\bibfnamefont {L.}~\bibnamefont {Dinis}},
  \bibinfo {author} {\bibfnamefont {D.}~\bibnamefont {Petrov}}, \ and\ \bibinfo
  {author} {\bibfnamefont {R.~A.}\ \bibnamefont {Rica}},\ }\href {\doibase
  10.1103/PhysRevLett.114.120601} {\bibfield  {journal} {\bibinfo  {journal}
  {Phys. Rev. Lett.}\ }\textbf {\bibinfo {volume} {114}},\ \bibinfo {pages}
  {120601} (\bibinfo {year} {2015})}\BibitemShut {NoStop}%
\bibitem [{\citenamefont {Arold}\ \emph {et~al.}(2017)\citenamefont {Arold},
  \citenamefont {Dechant},\ and\ \citenamefont {Lutz}}]{Arold2017}%
  \BibitemOpen
  \bibfield  {author} {\bibinfo {author} {\bibfnamefont {D.}~\bibnamefont
  {Arold}}, \bibinfo {author} {\bibfnamefont {A.}~\bibnamefont {Dechant}}, \
  and\ \bibinfo {author} {\bibfnamefont {E.}~\bibnamefont {Lutz}},\ }\href@noop
  {} {\bibfield  {journal} {\bibinfo  {journal} {arXiv:1707.06441}\ } (\bibinfo
  {year} {2017})}\BibitemShut {NoStop}%
\end{thebibliography}%
\end{document}